\documentclass[referee,sn-basic]{sn-jnl}

\usepackage{graphicx}%
\usepackage{multirow}%
\usepackage{amsmath,amssymb,amsfonts}%
\usepackage{amsthm}%
\usepackage{mathrsfs}%
\usepackage[title]{appendix}%
\usepackage{xcolor}%
\usepackage{textcomp}%
\usepackage{manyfoot}%
\usepackage{booktabs}%
\usepackage{algorithm}%
\usepackage{algorithmicx}%
\usepackage{algpseudocode}%
\usepackage{listings}%

\theoremstyle{thmstyleone}%
\theoremstyle{thmstyletwo}%

\theoremstyle{thmstylethree}%

\begin{document}

\title[Nearly homogeneous and isotropic turbulence generated by the interaction of supersonic jets]{Nearly homogeneous and isotropic turbulence generated by the interaction of supersonic jets}


\author[1]{\fnm{Takahiro} \sur{Mori}}

\author*[2]{\fnm{Tomoaki} \sur{Watanabe}}\email{watanabe.tomoaki@c.nagoya-u.jp}

\author[3]{\fnm{Koji} \sur{Nagata}}

\affil[1]{
\orgdiv{Department of Aerospace Engineering}, 
\orgname{Nagoya University}, 
\orgaddress{\city{Nagoya}, \postcode{464-8603}, \state{Aichi}, \country{Japan}}}

\affil*[2]{
\orgdiv{Education and Research Center for Flight Engineering}, 
\orgname{Nagoya University}, 
\orgaddress{\city{Nagoya}, \postcode{464-8603}, \state{Aichi}, \country{Japan}}}

\affil[3]{
\orgdiv{Department of Mechanical Engineering and Science}, 
\orgname{Kyoto University}, 
\orgaddress{\city{Kyoto}, \postcode{615-8530}, \state{Kyoto}, \country{Japan}}}


\abstract{
This study reports the development and characterization of a multiple-supersonic-jet wind tunnel designed to investigate the decay of nearly homogeneous and isotropic turbulence in a compressible regime. The interaction of 36 supersonic jets generates turbulence that decays in the streamwise direction. The velocity field is measured with particle image velocimetry by seeding tracer particles with ethanol condensation. Various velocity statistics are evaluated to diagnose decaying turbulence generated by the supersonic jet interaction. The flow is initially inhomogeneous and anisotropic and possesses intermittent large-scale velocity fluctuations. The flow evolves into a statistically homogeneous and isotropic state as the mean velocity profile becomes uniform. In the nearly homogeneous and isotropic region, the ratio of root-mean-squared velocity fluctuations in the streamwise and vertical directions is about 1.08, the longitudinal integral scales are also similar in these directions, and the large-scale intermittency becomes insignificant. The turbulent kinetic energy per unit mass decays according to a power law with an exponent of about 2, larger than those reported for incompressible grid turbulence. The energy spectra in the inertial subrange agree well with other turbulent flows when normalized by the dissipation rate and kinematic viscosity. The non-dimensional dissipation rate is within a range of 0.51--0.87, which is also consistent with incompressible grid turbulence. These results demonstrate that the multiple-supersonic-jet wind tunnel is helpful in the investigation of decaying homogeneous isotropic turbulence whose generation process is strongly influenced by fluid compressibility. 
}


\footnotetext[1]{\textcolor{red}{Preprint submitted to Experiments in Fluids.} }

\maketitle

\section{Introduction}\label{Sec_Introduction}
Compressible turbulence plays a crucial role in flows in engineering and physics fields~\citep{anderson1990modern,canuto1997compressible}. For sufficiently large velocity fluctuations, turbulent motion begins to cause fluid compression and expansion, by which fluid density varies. The compressibility effects significantly alter the evolution of turbulence. The growth rate of turbulent shear flows is suppressed at a high Mach number~\citep{bradshaw1977compressible}. When a Mach number defined for velocity fluctuations is very high, turbulent motion also locally generates shock waves called shocklets~\citep{lee1991eddy}. Velocity, pressure, density, and temperature drastically vary across shock waves, and the propagation of shocklets also influences turbulence. 
\par
Theoretical studies of turbulence often consider canonical turbulent flows under relatively simple conditions. These studies also mainly concern statistical behaviors of turbulence because of the difficulty in constructing theories and models for an instantaneous flow field based on the Navier--Stokes equations. One of the simplest problems is homogeneous isotropic turbulence (HIT)~\citep{batchelor1953theory}, for which various theories and models have been developed in previous studies, e.g., non-dimensional energy dissipation rate~\citep{lohse1994crossover}, longitudinal structure functions~\citep{bos2012reynolds}, and decay in absence of external force~\citep{davidson2009role}. Laboratory experiments are as crucial as numerical simulations in providing data that support the construction and validation of theories and models. Previous studies have developed various facilities and methods for generating HIT in a subsonic regime for which the assumption of incompressibility is valid. A grid installed in a uniform mean flow in a wind tunnel can generate nearly homogeneous and isotropic turbulence~(e.g., \citealt{comte1966use,uberoi1967effect,valente2011decay,kitamura2014invariants,djenidi2015power}). Similarly, experiments with water flumes utilize a grid to generate HIT~\citep{tan1963final,suzuki2010high}. Most grids consist of many intersecting bars, which stir a uniform mean flow. Another type of turbulence-generating grid is an active grid~\citep{makita1991realization}, which utilizes rotating bars with winglets. The random motion of winglets generates HIT with a high Reynolds number~\citep{mydlarski1996onset,larssen2011generation,zheng2021turbulent}. A similar method to randomly stir fluid is utilized in a multi-fan wind tunnel, by which the interaction of fan-induced flows generates HIT~\citep{ozono2018realization,takamure2019relative}. A mean flow advects the HIT generated in wind tunnels and water flumes. Because of negligible shear in the mean flow, turbulence decays as it evolves. The decay of grid turbulence is often compared with the theories of freely evolving HIT~\citep{krogstad2010grid}. \par
Previous studies have also developed facilities to generate turbulence without a mean flow. They often utilize stirring devices in a closed chamber. \cite{birouk2003attempt} reported the generation of HIT with electrical fans placed inside a chamber. Other studies also developed similar multi-fan facilities~\citep{semenov1965measurement,zimmermann2010lagrangian,ravi2013analysis,xu2017estimation,bradley2019measurement,yamamoto2022experimental}. Synthetic jet actuators are also widely used to generate HIT by the jet interaction~\citep{hwang2004creating,variano2004random}. Similarly, the interaction of random jets can generate HIT: \cite{bellani2014homogeneity} and \cite{carter2016generating} used two opposing planar arrays of randomly actuated jets while \cite{perez2016effect} used a single array. These facilities were used to investigate various phenomena related to turbulent flows, such as particle disperson~ \citep{hwang2006homogeneous} and evaporation of droplets~\citep{marie2017digital}. \par
Fewer experimental studies have been reported for HIT in compressible flows than in incompressible flows. Experiments of compressible turbulence have been extensively conducted for turbulent shear flows, such as jets~\citep{karthick2017passive}, mixing layers~\citep{goebel1990mean}, and wakes~\citep{bonnet1986large}, which have significant application in aerospace engineering. Compressible grid turbulence has been investigated in shock tube facilities. Grid turbulence is generated when the mean flow induced by a planar shock wave passes a grid installed inside a shock tube. Most studies of grid turbulence in shock tubes focus on the interaction between the shock wave and grid turbulence~\citep{honkan1992rapid,agui2005studies,fukushima2021impacts}. Exceptionally, \cite{briassulis2001structure} conducted comprehensive velocity measurements of grid turbulence in a shock tube. They compared the statistics of velocity fluctuations for various flow parameters, such as the Mach number of the mean flow and the Reynolds number based on the mesh size. The turbulent kinetic energy in grid turbulence is known to decay according to a power law given by $a_k(x-x_0)^{-n}$, with a coefficient $a_k$, a virtual origin $x_0$, and a decay exponent $n$. The experiments by \cite{briassulis2001structure} suggest that compressibility effects cause the decrease of $n$. 
Later, \cite{fukushima2021impacts} also observed the decrease of $n$ in compressible grid turbulence in a shock tube. \cite{zwart1997grid} reported compressible grid turbulence generated with a perforated plate installed in a high-speed wind tunnel. They found that $n$ in compressible grid turbulence exceeds values in incompressible cases when the Mach number is high. These results imply that the compressibility effects on grid turbulence are not universal and may depend on various factors, including the generation process of turbulence, as also discussed in \cite{zwart1997grid}. Indeed, even in an incompressible fluid, different decay laws of HIT were theoretically derived for the decay of Saffman turbulence and Batchelor turbulence, which differ in the shape of an energy spectrum at a low wavenumber range~\citep{davidson2004turbulence}. The low-wavenumber spectral shape depends on the turbulence generation process. Therefore, different compressibility effects are possible for the decay of HIT generated in different facilities. \par
Recent studies have developed other facilities for compressible turbulence. 
\cite{yamamoto2022turbulence} developed a compressible-turbulence chamber with opposing arrays of piston-driven synthetic jet actuators. The piston-driven actuators, initially proposed by \cite{crittenden2006high}, are capable of generating supersonic synthetic jets at a sufficiently high operation frequency~\citep{traub2012evaluation,sakakibara2018supersonic,tung2023large}. The interaction of supersonic synthetic jets generates turbulence strongly influenced by compressibility effects. Another recently developed facility for compressible turbulence is the variable density and speed-of-sound vessel. The experiments use a heavy gas (sulfur hexafluoride SF$_6$) as a working fluid. Even though a stirring device generates a flow with a velocity lower than 100~m/s, the Mach number can be high because of the low speed of sound of SF$_6$. These new facilities will be helpful in the future investigation of compressible turbulence. \par
As also found for the decay exponent of grid turbulence discussed above, flow-dependent compressibility effects on turbulence have been reported for other velocity statistics. In fully-developed incompressible turbulence, the skewness and flatness of a longitudinal velocity gradient, e.g., $\partial u/\partial x$, have universal relations with a turbulent Reynolds number~\citep{sreenivasan1997phenomenology}. Compressible turbulence tends to have larger absolute values of skewness and flatness than incompressible turbulence with a comparable Reynolds number~\citep{donzis2020universality,watanabe2021solenoidal}. One of the critical parameters in compressible turbulence is the turbulent Mach number, defined as $M_T=\sqrt{u_{rms}^2+v_{rms}^2+w_{rms}^2}/a$, where 
$u_{rms}$, 
$v_{rms}$, and 
$w_{rms}$ are the root-mean-squared (rms) velocity fluctuations in three directions, and $a$ is the speed of sound. $M_T$ is often treated as a dominant parameter determining compressibility effects~\citep{lee1991eddy,ristorcelli1997consistent}. However, a comparison of different compressible turbulent flows in experiments and direct numerical simulations (DNS) has shown that the compressibility effects on the velocity gradient statistics are not solely determined by local values of $M_T$~\citep{yamamoto2022experimental}: the deviation of the velocity derivative flatness from incompressible values occurs in supersonic turbulent jets and HIT generated by supersonic synthetic jets at much lower $M_T$ than in compressible HIT sustained by a solenoidal forcing. The dilatational fluid motion causes additional kinetic energy dissipation in strongly compressible turbulence. The dilational and solenoidal dissipation ratio is not also determined solely by $M_T$~\citep{donzis2020universality}. These behaviors are possibly explained by a non-local property of wave motion~\citep{yamamoto2022experimental}: even if a particular location in turbulence has low $M_T$, much higher values of $M_T$ can be observed in other regions, where strong pressure waves, such as shock waves, are generated and propagate into a flow region of interest. These studies imply that the compressibility effects on turbulence significantly depend on the generation process of turbulence. Investigating compressible turbulence in various facilities is crucial in extending our knowledge of turbulence behavior under compressibility effects. The further development of compressible turbulence facilities is demanded for this purpose. \par
The present study develops a new wind tunnel facility to investigate compressible decaying HIT. In this wind tunnel, the interaction of many supersonic jets generates nearly homogeneous and isotropic turbulence, which continuously decays in a test section without significant influences of mean shear. Velocity measurements with particle image velocimetry (PIV) and temperature and pressure measurements characterize the turbulent flow generated in the multiple-supersonic-jet wind tunnel, which is useful for investigating decaying HIT in a compressible regime. The paper is organized as follows. Section~\ref{sec_exp} describes the multiple-supersonic-jet wind tunnel and measurement methods. Section~\ref{sec_res} presents the measurement results to discuss the fundamental turbulence statistics compared with other turbulent flows. Finally, the paper is summarized in Sect.~\ref{Sec_Conclusions}. \par
\section{Experimental setup and measurement methods}\label{sec_exp}
\begin{figure}
 \begin{center}
  \includegraphics[width=1\linewidth, keepaspectratio]{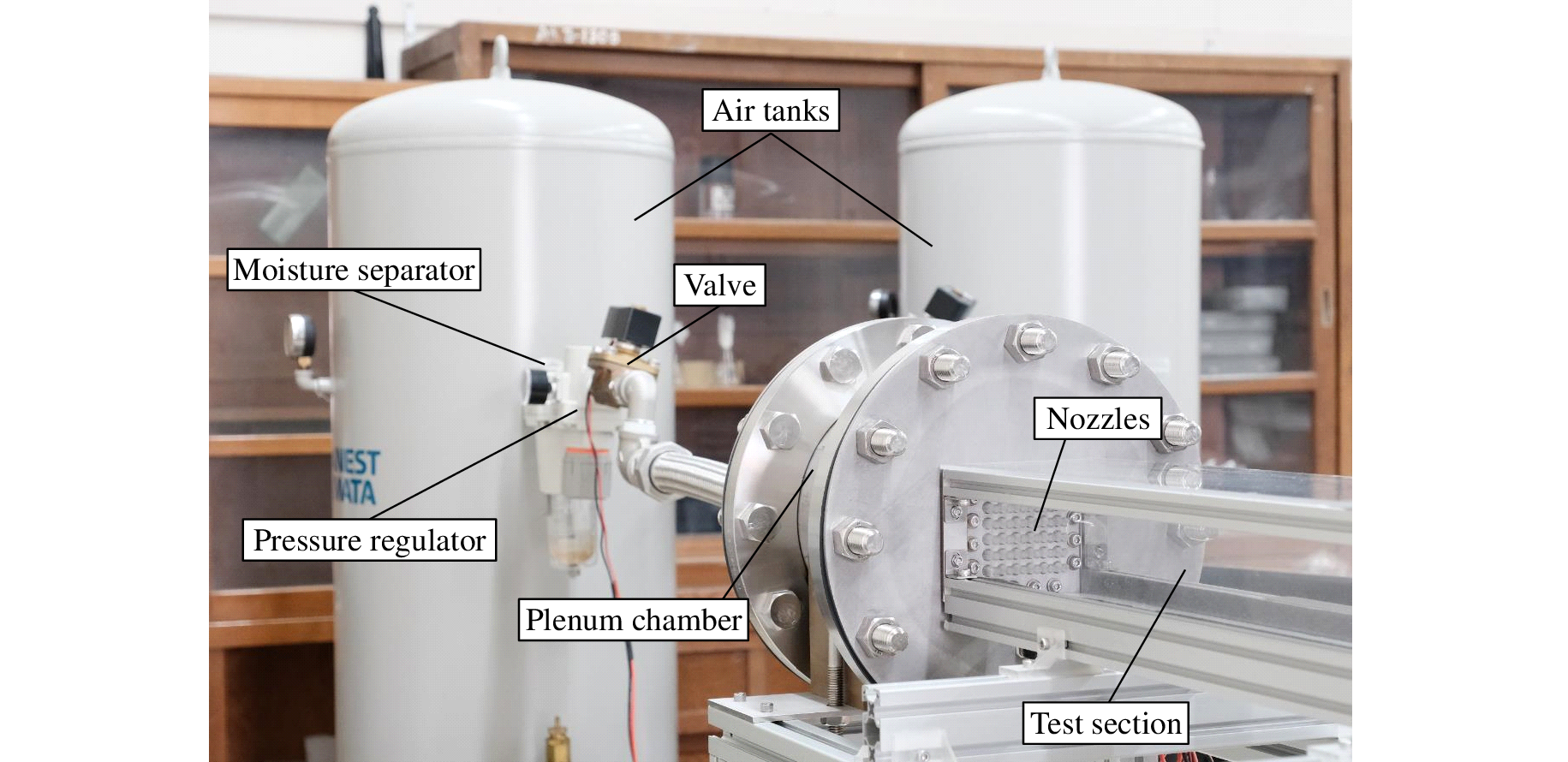}
  \caption{A multiple-supersonic-jet wind tunnel. }  
  \label{Fig_Pic}
 \end{center}
\end{figure}
\begin{figure}
 \begin{center}
  \includegraphics[width=1\linewidth, keepaspectratio]{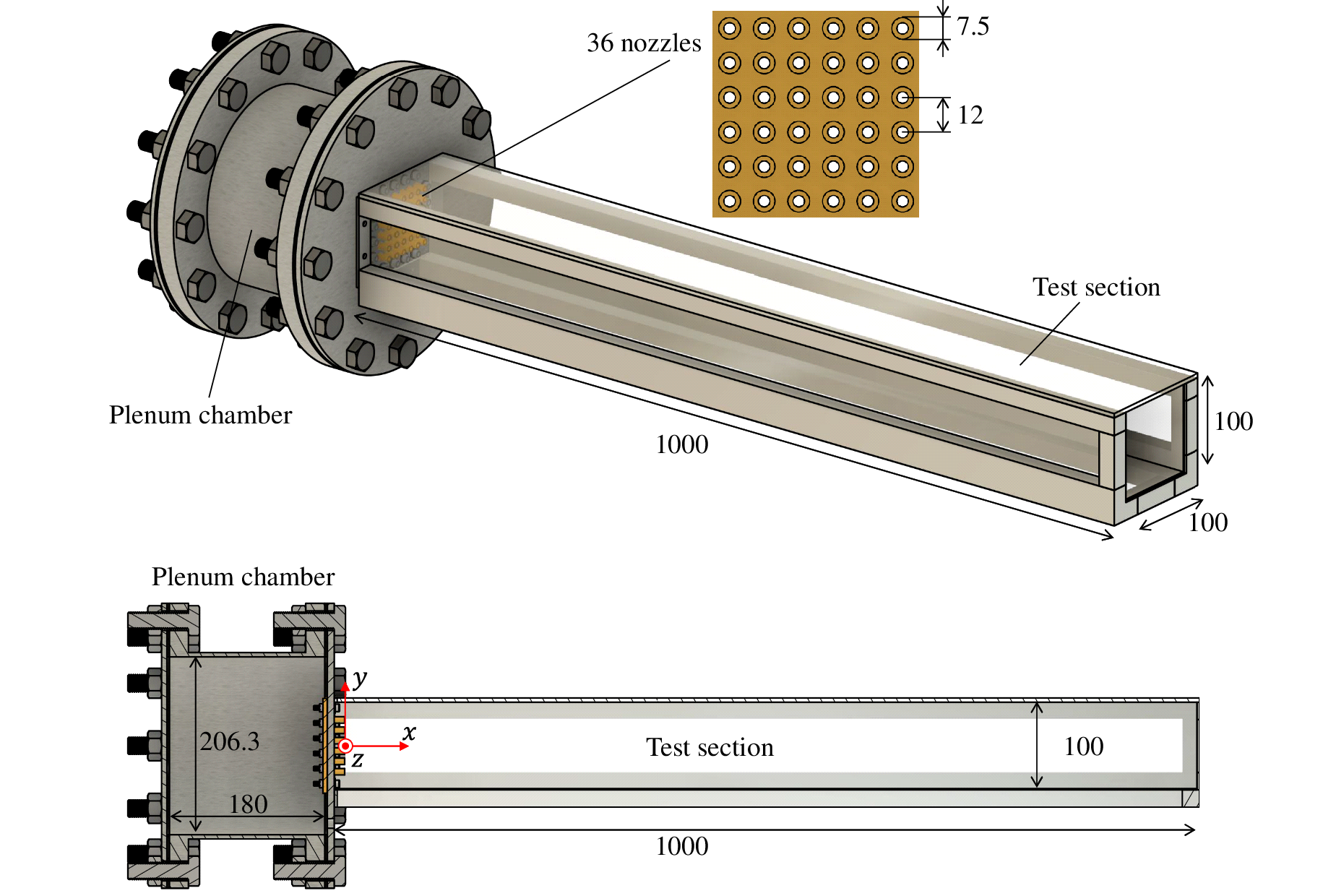}
  \caption{A schematic of a multiple-supersonic-jet wind tunnel. All dimensions
are in mm. }  
  \label{Fig_Schematic}
 \end{center}
\end{figure}
\subsection{Multiple-supersonic-jet wind tunnel}\label{sec_msjwt}
The multiple-supersonic-jet wind tunnel generates nearly homogeneous and isotropic turbulence by the interaction of supersonic round jets. The development of the facility is motivated by the experiments of incompressible HIT generated by the interaction of many subsonic jets~\citep{bellani2014homogeneity,carter2016generating,perez2016effect}. Figure~\ref{Fig_Pic} shows a picture of the wind tunnel with three main components: an air supply system, a plenum chamber, and a test section. Compressed air is supplied to the cylinder-shaped plenum chamber. The front surface of the plenum chamber is equipped with 36 Laval nozzles, which generate supersonic round jets. The interaction of 36 jets generates turbulence which eventually decays in the streamwise direction in the test section. This section describes the details of the wind tunnel. \par
Compressed air is stored in two air tanks with a volume of 220~L (Anest Iwata, SAT-220C-140). Compressors (Fujiwara Industrial Co., Ltd., SW-L30LPF-01) are connected to each tank via a moisture separator (SMC, AFM40-02-2A). The outlet port of each tank is connected to the back surface of the cylinder-shaped plenum chamber by a flexible hose with an inner diameter of 29.5~mm. Here, the air is supplied to the chamber through a moisture separator (SMC, AF60-10-A), a pressure regulator (SMC, AR60-10BG-B), and a pilot kick 2-port solenoid valve (CKD, ADK11-25A). The pressure regulator is used to adjust the plenum pressure. 
\par
\begin{figure}
 \begin{center}
  \includegraphics[width=1\linewidth, keepaspectratio]{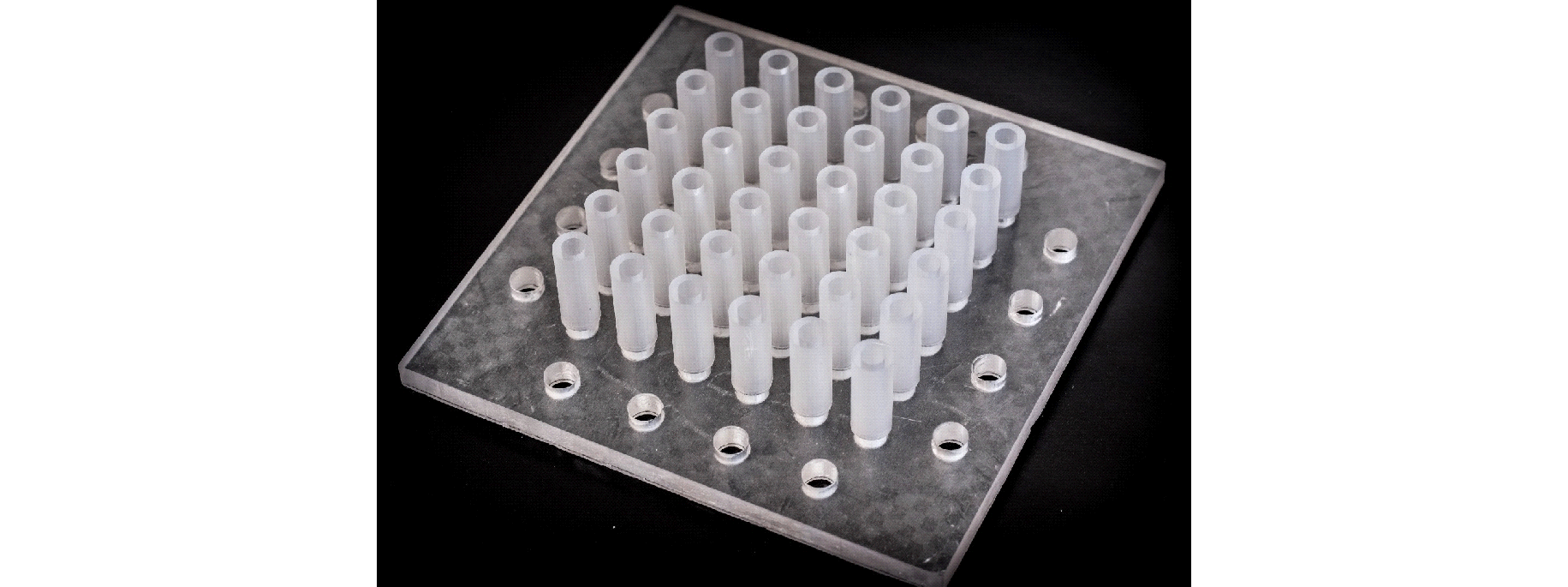}
  \caption{A nozzle plate with 36 Laval nozzles.  
}  
  \label{Fig_Pic_nozzles}
 \end{center}
\end{figure}
Figure~\ref{Fig_Schematic} shows a schematic of the plenum chamber and test section. The plenum chamber consists of a 180~mm length pipe and two round plates with a 330~mm diameter fixed on flanges welded to the pipe. The pipe size is JIS 20A, with an inner diameter of 206.3~mm and a thickness of 5~mm. The pipe, plates, and flanges are made of stainless steel. Rubber seals designed for JIS 20A pipes fill the space between the plates and flanges. The flexible hoses are connected to two threaded holes on the back plate with a thickness of 6~mm. A pressure sensor (SMC, ISE30A-C6L-N-M) connected to a port on the front plate monitors the pressure in the plenum chamber. The plenum pressure is also measured with another pressure sensor (SMC, PSE-540A R06), whose signal is recorded by an oscilloscope (Yokokawa, DL850E) at a sampling rate of 1~kHz. \par
A nozzle plate with $6\times6$ Laval nozzles is installed on the front plate. Figure~\ref{Fig_Pic_nozzles} shows a picture of the nozzle plate manufactured with stereolithography 3D printing with a printing resolution of 10~$\mu$m. All 36 nozzles have the same geometry. The holes along the perimeter of the nozzle plate are used to fix the plate to the plenum chamber. The inner diameters at the jet exit and the throat are 4.31~mm and 4.12~mm, respectively. The nozzle inlet diameter is 6.3~mm. The constant outer diameter is 7.5~mm. The center-to-center distance between the two nozzles is 12~mm. The inner diameter of the converging section varies according to a square of a sinusoidal function. The internal geometry of the diverging section is determined with an analytical method for nozzle design~\citep{foelsch1949analytical}. Each nozzle with these dimensions generates a fully expanded supersonic jet with a Mach number of 1.36 in atmospheric air when the gauge pressure in the chamber is 200~kPaG. 
The base of the nozzle plate has a square shape of $110\times110$~mm$^{2}$ and a thickness of 5~mm. The front plate of the chamber has holes with a diameter of 9~mm at the locations corresponding to the nozzles. The nozzles are inserted and fixed from the inner side of the chamber plate with bolts and nuts. Air leakage is prevented with seal washers used with the nuts and a hand-cut PTFE gasket sheet with a 1~mm thickness, which fills the space between the nozzle plate and the chamber plate. \par
The test section has $100\times100$~mm$^2$ square cross-section and 1000~mm length. The test section is made of an aluminum frame and acrylic plates. For the side and bottom walls, optical-grade acrylic sheets (Nitto Jushi Kogyo, CLAREX) with a 1.5~mm thickness are used for optical measurements and visualizations. The top wall is a conventional acrylic plate with a 3~mm thickness. The plenum chamber is placed on a urethane pipe-holder designed for JIS 20A pipes (Nichiei Intec, N-040219). The test section and pipe holder are fixed on another aluminum frame. \par
\begin{figure}
 \begin{center}
  \includegraphics[width=1\linewidth, keepaspectratio]{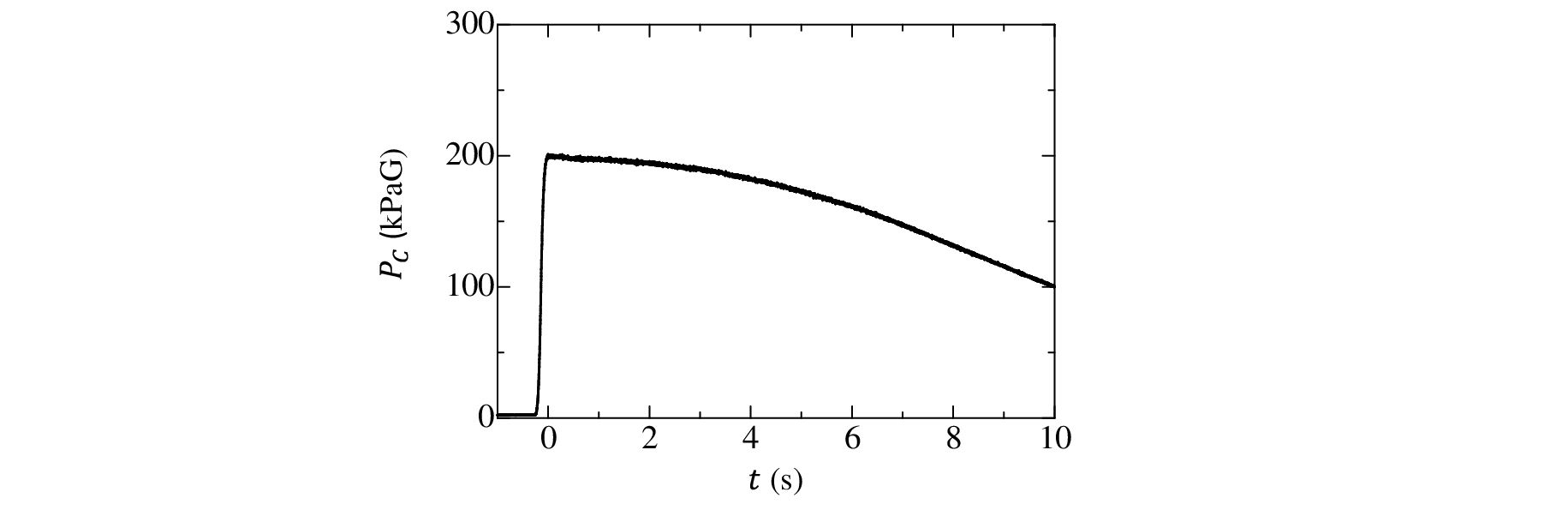}
  \caption{Temporal variation of gauge pressure in the plenum chamber, $P_C$. $t=0$ corresponds to the maximum value of $P_C$. 
}  
  \label{Fig_PC}
 \end{center}
\end{figure}
All experiments are conducted to generate the supersonic jets with a plenum pressure of $200$~kPaG. In each run of the experiments, the air tanks are filled with compressed air with a gauge pressure of 900~kPaG. Once the valves are opened, the compressed air flows into the plenum chamber, increasing the plenum pressure. The pressure regulators between the air tanks and the plenum chamber are adjusted to realize the desired value of the plenum pressure. Figure~\ref{Fig_PC} presents the temporal variation of the gauge pressure in the plenum chamber, $P_C$, during the experiment. In this plot, $t=0$ corresponds to the maximum value of $P_C$. After the valve is opened, the pressure immediately increases to 200~kPaG. The pressure hardly varies with time for about 3~s, and then begins to decrease because the pressure in the air tanks decreases. The plenum pressure decreases from 200~kPaG by 5\% at $t=2.9$~s. The flow in the test section is statistically steady for about 2~s, for which the measurements are conducted in this study. 
The streamwise, vertical, and spanwise directions are denoted by $x$, $y$, and $z$, respectively, and the origin of the coordinate system is at the center of the jet exits of the nozzle plate, as shown in Fig.~\ref{Fig_Schematic}. \par
\subsection{Particle image velocimetry}
Velocity measurements are conducted with two-dimensional and two-component PIV. The present study uses the DANTEC PIV system, which consists of a double-pulse Nd:YAG laser (Dantec Dynamics, Dual Power 65-15) and a high-speed camera (Dantec Dynamics, SpeedSense 9070) with a camera lens with a focal length of 105~mm (Nikkon, AI AF Micro Nikkor 105mm F2.8D). The system is controlled with a synchronizer (Dantec Dynamics, 80N77) and PIV software (Dantec Dynamics, Dynamic Studio). Light sheet optics mounted on the laser produce a thin light sheet with a thickness smaller than 1~mm. The laser unit is placed under the wind tunnel, and a mirror in front of the laser sheet optics is used to illuminate an $x$--$y$ plane from the bottom of the test section. The camera captures particle images from the side of the test section. The time interval of two laser pulses is adjusted depending on the measurement location, as discussed below. Each laser pulse has a width of less than 4~ns. The particle-pair images are captured at 15~Hz. The images are processed with an adaptive PIV algorithm~\citep{theunissen2010adaptive} and universal outlier detection~\citep{westerweel2005universal} implemented in Dynamic Studio. \par
The present PIV uses condensed ethanol droplets generated by the Laval nozzles as tracer particles. Liquid ethanol is seeded and evaporated in the air tanks before the experiments. When fluid with evaporated ethanol passes the Laval nozzles, fluid expansion in the diverging sections causes a temperature drop, resulting in the condensation of ethanol. This seeding technique has widely been used in supersonic wind tunnels, which also use divergent nozzles to produce supersonic flows. Previous studies have confirmed that generated particles have diameters less than 1~$\mu$m: \cite{pizzaia2018effect} reported that the diameter of ethanol droplets was approximately $0.05$--0.2~$\mu$m; \cite{kouchi2019acetone} used acetone droplets produced by a Lavar nozzle for PIV measurement, by which the droplet diameter was estimated as 160~nm. These studies have shown that the condensation droplets are small enough for PIV measurements of supersonic flows. The diameter can be estimated from light scattered by droplets. By visibly observing particles illuminated by white LED (Sigmakoki, SLA-100A), the droplet diameter is confirmed to be in the Rayleigh-scattering regime and smaller than visible light wavelength, 360--830~nm. The present study investigates turbulence generated by the interaction of supersonic jets. As the jets mix with ambient fluid, the mean velocity in the measurement area is in a subsonic regime. \cite{kouchi2019acetone} conducted accurate PIV measurements of supersonic flows with condensation particles. Therefore, for the present subsonic velocity range, the ethanol droplets are useful as tracer particles of PIV. \par
Because how the ethanol is injected in the air tanks affects the number density of tracer particles in the test section, we describe the detailed procedure of the experiments. The air tanks have three ports: an inlet port connected to the compressor, an outlet port connected to the plenum chamber, and an exhaust port used to release compressed air after experiments. The exhaust port can be opened or closed by a hand-operated valve, while the inlet port can be opened by disconnecting a gas hose from the compressor. The following procedure is repeated for the two air tanks. First, ethanol is sprayed into the air tank via the inlet port by using an airbrush. Sprayed ethanol quickly evaporates in the tank. When ethanol is injected, the exhaust valve is kept partially opened to prevent the reversal flow from the inlet port. 70~ml of liquid ethanol is injected into each tank. The exact amount of ethanol in the tank is difficult to estimate because the air partially leaks from the tank while spraying the ethanol. However, we have confirmed that increasing the amount from 70~ml does not influence particle images captured by the PIV system, whereas smaller amounts result in fewer particles. After the compressor is connected to the inlet port of the tank again and the exhaust valve is closed, the air is stored in the tank until the gauge pressure reaches 900~kPaG. Even for a supersaturated case, the mass fraction of ethanol at this stage is smaller than 1\%, for which the ethanol/air mixture and condensation are expected to have negligible influences on the turbulent flow in the test section~\citep{clemens1991planar,pizzaia2018effect}. We have also confirmed that ethanol does not change the average and standard deviation of dynamic pressure measured with a Pitot tube. Upon opening the solenoidal valves between the tanks and the plenum chamber, the camera starts capturing particle images. Velocity vectors are calculated by analyzing the images captured while the plenum pressure is constant. \par
\begin{table}
 \begin{center}
 \caption{Locations and parameters of PIV measurements. }
 \label{Table_PIV}
 \begingroup
 \begin{tabular}{cccccccccccc}
 \hline
  $x_C$
  & 0.150 m
  & 0.267 m
  & 0.360 m
  & 0.465 m
  & 0.605 m
  & 0.746 m
  \\
  $y_C$
  & 0 m
  & 0 m
  & 0 m
  & 0 m
  & 0 m
  & 0 m
  \\
  $\Delta t$ 
  & 3 $\mu$s
  & 4 $\mu$s
  & 4 $\mu$s
  & 4 $\mu$s
  & 4 $\mu$s
  & 4 $\mu$s
  \\
  $N_S$ 
  & 722 
  & 797 
  & 700 
  & 726 
  & 651 
  & 595 

  \\
 \hline
 \end{tabular}
 \endgroup
 \end{center}
\end{table}
The PIV measurements are conducted at six streamwise locations. Table~\ref{Table_PIV} summarizes the measurement locations and parameters. The center of the measurement area is $(x,y)=(x_C, y_C)$. These locations are chosen to investigate turbulence generated by the jet interaction and do not include the region where each jet evolves before the interaction. This is because our primary interest is in a nearly homogeneous and isotropic region where turbulence decays. Although the supersonic jets are issued in the test section, the mean velocity in the measurement locations is subsonic because the high-speed jets mix with surrounding fluids with a small mean velocity. The table also shows the time interval $\Delta t$ between two pulses of the PIV. At $x_C=0.150$~m, the flow is highly inhomogeneous in the cross-section, and the mean velocity along the centerline is higher than the surroundings. Therefore, $\Delta t$ is smaller at $x_C=0.150$~m than in the downstream region. These $\Delta t$ values are determined based on the particle displacement during $\Delta t$. The spatial resolution, defined as the spacing between points to calculate velocity vectors, is about 1~mm in both directions. As shown below, the present PIV resolves the scales close to the small-scale end of the inertial subrange and does not resolve the smallest scale of turbulent motion. This resolution is sufficient to evaluate most velocity statistics dominated by large-scale motion. In each wind tunnel operation, about 35 snapshots of velocity vectors are captured on average. The experiments are repeated many times for statistical analysis. The total number of acquired vector images, $N_S$, is also shown in Table~\ref{Table_PIV}. The snapshots of velocity vectors are measured at $15$~Hz, for which the time interval between two snapshots is $0.067$~s. The mean velocity in the fully developed turbulent region is about 40~m/s. The characteristic length scale of the flow is roughly estimated as the spacing between the nozzles, 0.012~m. The flow time scale defined with these velocity and length scales is $0.0003$~s, much shorter than the sampling interval. Thus, the snapshots of velocity vectors are independent samples that are not correlated with each other. Statistics are evaluated with ensemble averages. Two-dimensional velocity vectors are denoted by $(u,v)$. For a varibale $f$, the $n$th snapshot is denoted by $f^{(n)}(x,y)$. The average of $f$,  $\langle f\rangle$, is evaluated as
\begin{align}
  \langle f\rangle(x,y)=
  \frac{1}{N_S}\sum^{N_{S}}_{n=1}f^{(n)}(x,y).
\end{align}
Fluctuations are defined as ${f'}^{(n)}(x,y)=f^{(n)}(x,y)-\langle f\rangle(x,y)$, and the rms fluctuations are evaluated as $f_{rms}(x,y)=\langle {f'}^2\rangle^{1/2}=(\langle f^2\rangle-\langle f\rangle^2)^{1/2}$. In addition, skewness and flatness of $f$ are defined as 
$S(f)=\langle f'^3\rangle/\langle f'^2\rangle^{3/2}$ and 
$F(f)=\langle f'^4\rangle/\langle f'^2\rangle^{4}$, respectively. \par
\subsection{Pitot tube measurement}
Pitot tube measurements are conducted to investigate the flow evolution along the centerline of the test section. The present study uses an L-shaped standard Pitot tube with a 7~mm diameter (Testo, 0635 2145) and a pressure sensor (SMC, PSE543A-R06). The signal from the pressure sensor is recorded with an oscilloscope (Yokokawa, DL850E) at a sampling rate of 10~kHz. Total pressure $P_T$ is measured at the tip of the tube, while static pressure $P_S$ is measured on the side of the tube, which is separated from the tip by 30~mm in the streamwise direction. The top wall of the test section is replaced with many small plates between which the Pitot tube is inserted into the test section. The measurement location is along the centerline of the test section. As shown in Sect.~\ref{sec_press}, the streamwise distribution of static pressure $P_S(x)$ is well fitted by an exponential curve  $P_{S, F}(x)=a_P(b_P)^{x}$, where the parameters can be obtained with a least squares method for measured $P_S$. Because the Pitot tube measures $P_T(x)$ and $P_{S,F}(x+\Delta_{Pitot})$ with $\Delta_{Pitot}=30$~mm, the pressure difference $\Delta P(x)=P_T(x)-P_S(x)$ is calculated with the measured $P_T(x)$ and the fitting curve $P_{S,F}(x)$. A time average is calculated for the time series pressure data over 2~s, for which the plenum pressure is almost constant. \par
Because the measurements are conducted in turbulence, the velocity fluctuations also affect $\Delta P$, whose measured values are not related solely to the dynamic pressure due to the mean flow. Turbulence increases the total pressure measured by the Pitot tube. The following relation expresses the pressure difference measured by a Pitot tube in turbulence~\citep{bailey2013obtaining}:
\begin{align}
  \Delta P = \frac{1}{2}\rho\left( 
  \langle u\rangle^2 
 + \langle {u'}^2\rangle
 + \langle {v'}^2\rangle
 + \langle {w'}^2\rangle
\right). 
\label{Eq_Pitot}
\end{align}
The velocity variances are evaluated with the PIV as
  $\langle {u'}^2\rangle
 + \langle {v'}^2\rangle
 + \langle {w'}^2\rangle\approx\langle {u'}^2\rangle+2\langle {v'}^2\rangle$. 
Here, the Pitot tube measurement is conducted along the centerline, where $\langle {v'}^2\rangle\approx  \langle {w'}^2\rangle$ is expected to be valid. The density is also estimated from static pressure and temperature measurements, as described below. Thus, the mean velocity is calculated from $\Delta P$ measured by the Pitot tube as 
\begin{align}
  \langle u\rangle=
\sqrt{\frac{2\Delta P}{\rho} -(\langle {u'}^2\rangle+2\langle {v'}^2\rangle)}.
\label{Eq_PitotU}
\end{align}
\par
Another factor that may affect the Pitot tube measurement is the compressibility effect. However, based on the mean velocity measured with PIV, we have estimated that this effect results in underestimating the mean velocity only by less than 2\%. Therefore, the compressibility correction for mean velocity is not applied to the Pitot tube measurement. \par
\subsection{Temperature measurement}
Temperature is measured with a fine sheathed K-type thermocouple (J Thermo, TJK-LS1501GP), whose sheath has a 0.15~mm diameter and a 100~mm length. The response time of this thermocouple is less than 1~s. The output voltage of the thermocouple is processed and recorded by a data logger (Hioki, LR8431) at a sampling rate of 100~Hz. Because of the response time, the temperature recorded over 1~s after 1~s from the start of the wind tunnel is averaged to estimate the mean temperature. The thermocouple is inserted from the top of the test section in the same manner as the Pitot tube. Temperature measurements are conducted at different streamwise locations. \par
\subsection{Estimation of the dissipation rate of turbulent kinetic energy}\label{sec_dis}
The decay of turbulent kinetic energy along the centerline is investigated for a nearly homogeneous and isotropic region corresponding to the downstream region. The turbulent kinetic energy per unit mass, $k_T=(\langle {u'}^2\rangle
 + \langle {v'}^2\rangle
 + \langle {w'}^2\rangle)/2$ is evaluated as $(\langle {u'}^2\rangle+2\langle {v'}^2\rangle)/2$ with the PIV data. The governing equation for $k_T$ can be derived from the conservation equation of momentum for a compressible fluid, which is rewritten in the form of 
\begin{align}
   \frac{\partial u_{i}}{\partial t}
  +u_{j}\frac{\partial u_{i}}{\partial x_{j}}
   =-\frac{1}{\rho}\frac{\partial p}{\partial x_{i}}
   +\frac{1}{\rho}
    \frac{\partial \tau_{ij}}{\partial x_{j}}.
   \label{Eq_mome}
\end{align}
Here, $p(x,y,z,t)$ and $\rho(x,y,z,t)$ are local values of pressure and density, respectively, and $\tau_{ij}$ is the viscous stress tensor. Multiplying Eq.~(\ref{Eq_mome}) by $u'_{i}=u_{i}-\langle u_{i}\rangle$ and rearranging terms yield the following governing equation for $k_T$:
\begin{align}
   \frac{\partial k_T}{\partial t}
+\langle u_{j}\rangle\frac{\partial k_T}{\partial x_j}
=
\underbrace{
-\langle u'_{i}u'_{j}\rangle
 \frac{\partial \langle u_{i}\rangle}{\partial x_j}
}_{P_k}
\underbrace{
-\frac{1}{2}\frac{\partial \langle u'_{i}u'_{i}u'_{j}\rangle}{\partial x_j}
}_{D_T}
+
\underbrace{
\left\langle \frac{u'_{i}u'_{i}}{2}\frac{\partial u'_{j}}{\partial x_j}
\right\rangle
}_{\Theta_k}
\nonumber\\
\underbrace{
-\left\langle \frac{1}{\rho}\frac{\partial u'_{j} p}{\partial x_j}\right\rangle 
}_{D_P}
+
\underbrace{
\left\langle \frac{p}{\rho}\frac{\partial u'_{j}}{\partial x_j}\right\rangle 
}_{\Theta_P}
+
\underbrace{
\left\langle \frac{1}{\rho}\frac{\partial u'_{i} \tau_{ij}}{\partial x_j}\right\rangle
}_{D_V}
-
\underbrace{\left\langle \frac{\tau_{ij}}{\rho}\frac{\partial u'_{i} }{\partial x_j}\right\rangle
}_{\varepsilon}. 
  \label{Eq_kt}
\end{align}
The first and second terms of the left-hand side represent the temporal variation of $k_T$ and the advection due to mean flow, respectively. On the right-hand side, $P_k$ is the production term, $D_T$ is the turbulent diffusion term, $D_P$ is the pressure diffusion term, $D_V$ is the viscous diffusion term, and $\varepsilon$ is the dissipation term. These terms also appear in the governing equation for an incompressible fluid. The remaining terms represent direct influences of fluid compressibility: $\Theta_k$ is the change in the turbulent kinetic energy per unit mass due to fluid expansion or compression, and $\Theta_P$ is the energy conversion between the kinetic and internal energies by pressure work. 
Even though the interaction of the supersonic jets generates turbulence near the nozzles, the velocity fluctuations in the decay region are subsonic, implying that $\Theta_k$ and $\Theta_P$ are not dominant in the variation of $k_T$ once the nearly homogeneous and isotropic turbulence has fully developed. As long as the Reynolds number is not too small, the viscous contribution to the spatial transport of kinetic energy is negligible, and $D_V$ is much smaller than other terms in turbulence~\citep{watanabe2016large,wang2022interscale}. In addition, the flow is assumed to be statistically steady and homogeneous in the $y$ and $z$ directions, and the diffusion terms in these directions are negligible. The turbulent and pressure diffusions in the streamwise direction are also negligible compared with the transport due to the mean flow in decaying HIT~\citep{wang2022interscale}. Then, Eq.~(\ref{Eq_kt}) is simplified as  
\begin{align}
\langle u\rangle\frac{\partial k_T}{\partial x}
=-\varepsilon.
  \label{Eq_Epsilon}
\end{align}
This relation is often used to evaluate the dissipation rate in grid turbulence because $\langle u\rangle$ and $k_T$ are much easier to measure than $\varepsilon$~\citep{kistler1966grid,thormann2014decay}. The decay of HIT depends on the spectral shape at large scales, which can differ depending on how turbulence is generated~\citep{davidson2004turbulence}. The compressibility effect on the decay properties may also emerge through the generation process of turbulence, although this influence does not appear in the governing equation of $k_T$. Such influences may appear in the virtual origin and decay exponent in the power law decay of $k_T$~\citep{briassulis2001structure,zwart1997grid}. 
\par
Equation~(\ref{Eq_Epsilon}) is used to estimate the turbulent kinetic energy dissipation rate from the PIV data. The decay of $k_T$ in HIT is well represented by a power law given by 
\begin{align}
  k_T=a_k(x-x_0)^{-n},
\label{Eq_decay}
\end{align}
where $a_k$ is a coefficient related to the level of velocity fluctuations, $x_0$ is the virtual origin, and $n$ is the decay exponent~\citep{mohamed1990decay,davidson2004turbulence}. A least squares method is applied to the streamwise distribution of $k_T$ obtained by the PIV data. Then, $\varepsilon$ is evaluated as  
\begin{align}
\varepsilon
=\langle u\rangle an(x-x_0)^{-n-1}.
  \label{Eq_Epsilon2}
\end{align}
\begin{figure}
 \begin{center}
  \includegraphics[width=1\linewidth, keepaspectratio]{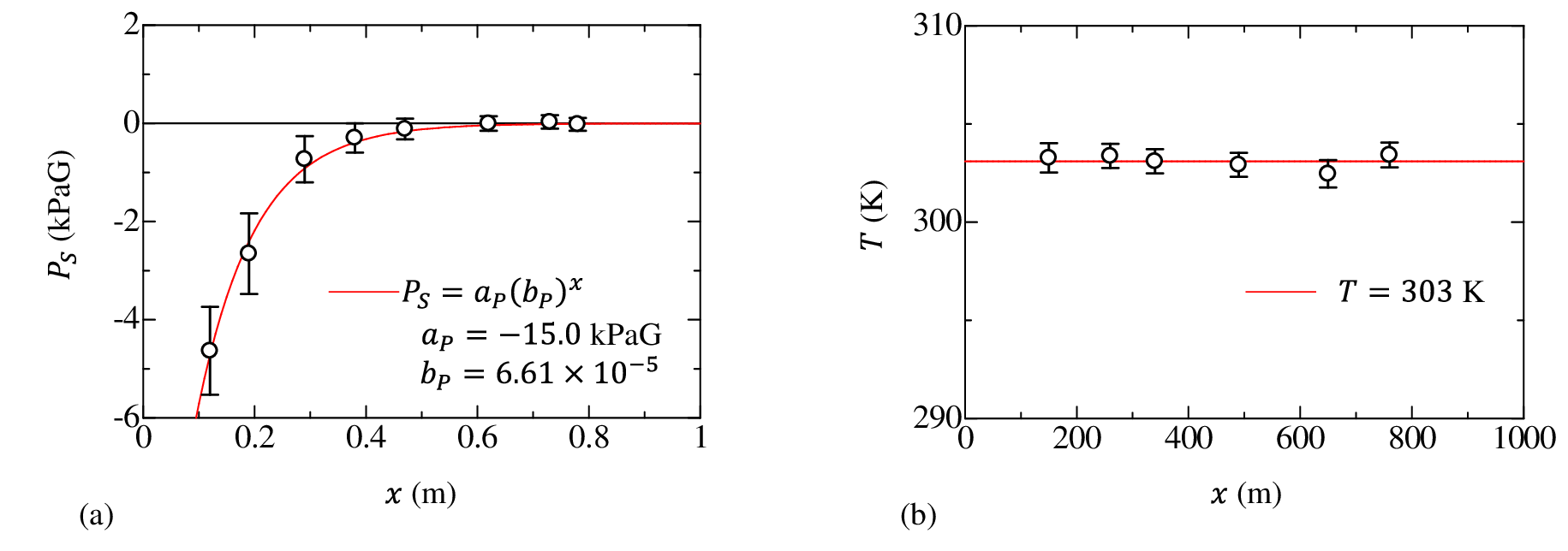}
  \caption{Streamwise variations of (a) mean static pressure $P_S$ and (b) mean temperature $T$. The error bars represent the standard deviation.  
}  
  \label{Fig_PsT}
 \end{center}
\end{figure}
\section{Results and discussion}\label{sec_res}
\subsection{Pressure and temperature profiles}\label{sec_press}
Figure~\ref{Fig_PsT}(a) presents the streamwise variation of the mean static pressure (gauge pressure), $P_S$. The error bars represent the standard deviations of time series pressure histories, which indicate the degree of temporal pressure fluctuations. The pressure increases in the downstream direction and approaches the atmospheric pressure. The pressure distribution is well approximated by an exponential curve $P_S=a_P(b_P)^x$, where $a_P=-15.0$~kPaG and $b_P=6.61\times 10^{-5}$ are obtained with a least squares method. As the results are shown as gauge pressure, this pressure variation is slight compared to the atmospheric pressure and hardly affects the mean fluid density in the test section. Figure~\ref{Fig_PsT}(b) presents the mean temperature $T$ along the centerline. The error bars also represent the standard deviation. The streamwise variation of $T$ is slight, and the average of all data points yields 303~K. \par
These measurement results enable us to estimate fluid density $\rho$ with the equation of state for an ideal gas as $\rho=P/RT$ with the gas constant $R=287.05$~J/kg K. Here, the exponential function $P_S=a_P(b_P)^x$ and mean temperature of 303~K are used to evaluate $\rho(x)$. The mean temperature yields the viscosity coefficient $\mu$, calculated with Sutherland's law as 
\begin{align}
  \mu = \mu_0\left(\frac{T}{T_0}\right)
\left(\frac{T_0+S}{T+S}\right),
\label{Eq_mu}
\end{align}
with $T_0=273$~K, $S=110.4$~K, and $\mu_0=1.724\times10^{-5}$ Pa s. As the mean temperature does not vary with $x$, $\mu$ is also treated as a constant. These fluid properties are used in the analysis of velocity statistics obtained with the PIV. 
 \par
\subsection{Velocity statistics}\label{sec_velocity}
\begin{figure}
 \begin{center}
  \includegraphics[width=1\linewidth, keepaspectratio]{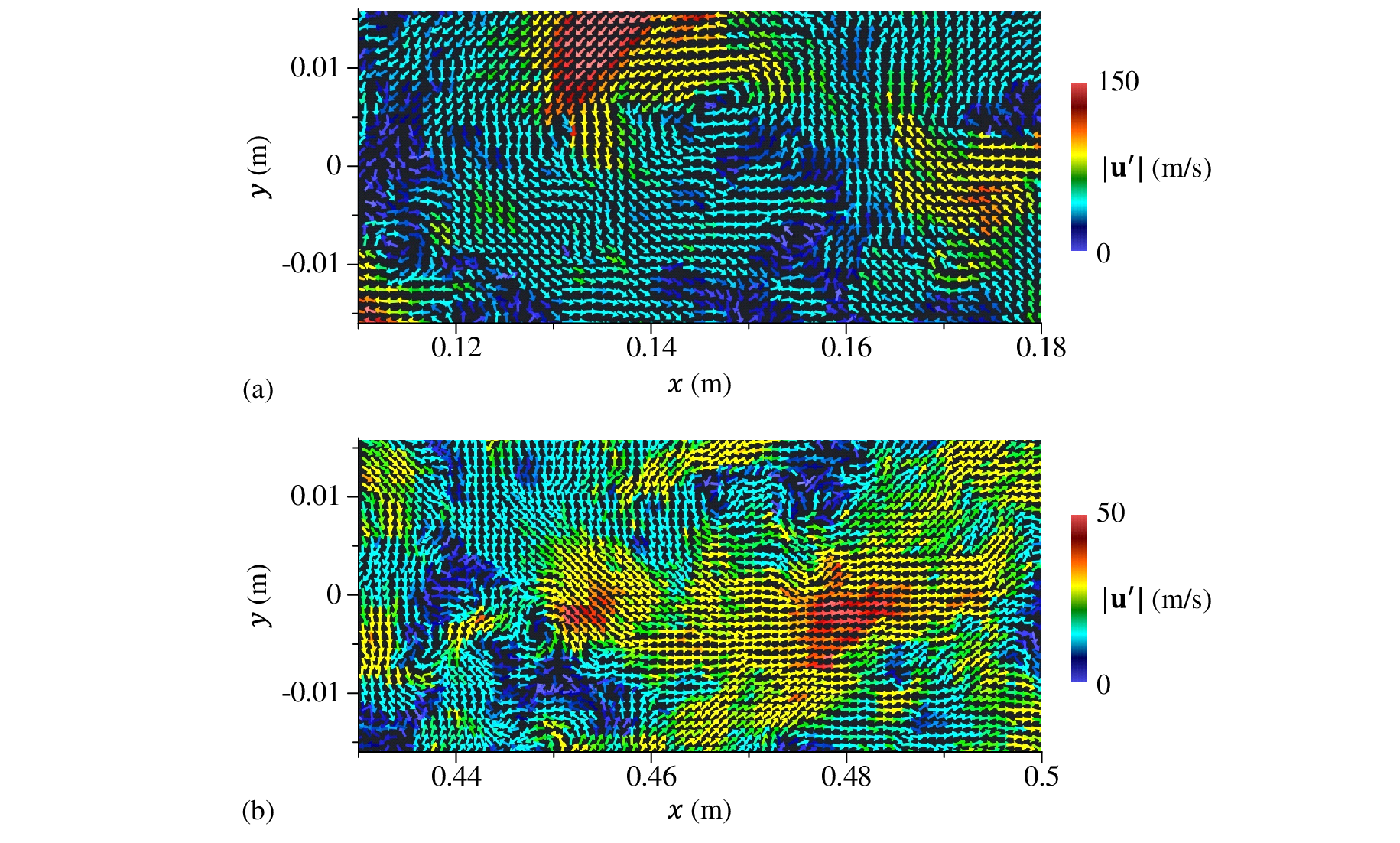}
  \caption{
Instantaneous fluctuating velocity vectors ${\bf{u}'}=(u',v')$ measured at (a) $x_C=0.150$~m and (b) $x_C=0.465$~m. The color represents the vector magnitude $|{\bf{u}'}|$. 
}  
  \label{Fig_UVinst}
 \end{center}
\end{figure}
Figure~\ref{Fig_UVinst} shows instantaneous fluctuating velocity vectors ${\bf{u}'}=(u',v')$ measured at $x_C=0.150$~m and $0.465$~m. The color represents the vector magnitude. For both measurement locations, the turbulent jets from the nozzles have fully developed and have interacted with each other, and the clear imprints of each jet are not observed in the velocity field. At $x_C=0.150$~m, the velocity fluctuations reach about 150~m/s. As turbulence decays, velocity fluctuations become small from $x_C=0.150$~m to $0.465$~m. At both locations, the vector fields exhibit typical rotating patterns associated with vortices, e.g., $(x,y)=(0.11$~m, $-0.01$~m$)$ in Fig.~\ref{Fig_UVinst}(a), indicating that the present PIV well captures these turbulent structures. \par
\begin{figure}
 \begin{center}
  \includegraphics[width=1\linewidth, keepaspectratio]{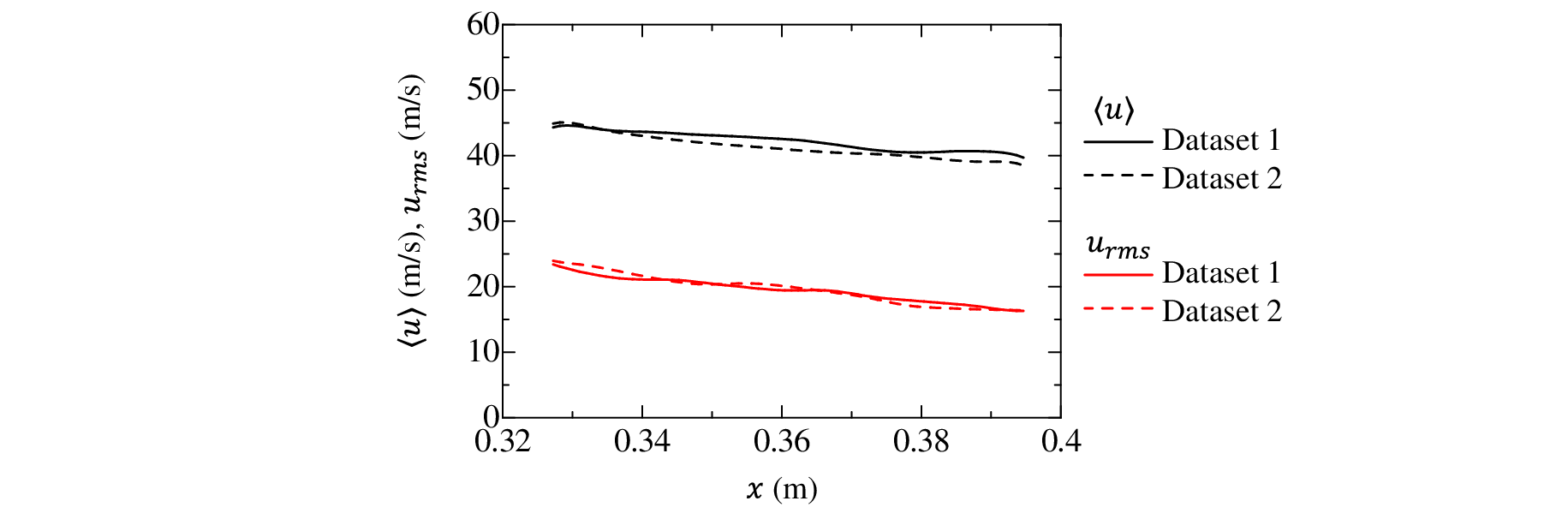}
  \caption{
Statistical convergence tests for mean velocity and rms velocity fluctuations with the PIV data measured at $(x_C, y_C)=(0.360$~m, $0$~m$)$. Totally 700 vector snapshots are divided into two datasets, each containing 350 snapshots. The mean velocity and rms velocity fluctuations calculated separately for the two datasets are plotted as functions of $x$ at $y=0$. 
}  
  \label{Fig_Convergence}
 \end{center}
\end{figure}
The degree of statistical convergence is examined before the velocity statistics are discussed. For the measurement location of $(x_C, y_C)=(0.360~$m, $0$~m$)$, 700 snapshots of velocity vectors are acquired. These snapshots are divided into two datasets, each containing 350 snapshots. The velocity statistics are separately calculated for the two datasets whose differences arise from statistical errors. Figure~\ref{Fig_Convergence} compares the streamwise distributions of the average and rms fluctuations of streamwise velocity, $\langle u\rangle$ and $u_{rms}$, at $y=0$. The two datasets yield slightly different distributions. A statistical quantity $f$ calculated with datasets 1 and 2 is denoted by $f_{1}$ and $f_{2}$, respectively. For the distribution of $f$ along the $x$ axis at $y=0$, the statistical error arising from the finite number of samples is evaluated with $f_{1}(x,0)$ and $f_{2}(x,0)$ as an rms error: 
\begin{align}
E(f)= \sqrt{\frac{1}{x_S-x_E}\int_{x_S}^{x_E} (f_{1}-f_{2})^{2}dx},
    \label{Eq_Err}
\end{align}
where $x_S\leq x \leq x_E$ is the measurement area. For Fig.~\ref{Fig_Convergence}, the relative errors, $E(\langle u\rangle)/\langle u\rangle$ and $E(u_{rms})/u_{rms}$, are about 5\%, which is a typical statistical error when the number of the snapshots is about 350. Thus, statistical errors for the present PIV measurement with $N_S>500$ are expected to be smaller than 5\%. The statistical errors of other velocity statistics are evaluated with this method for each PIV measurement location, and $E(f)$ is shown in figures below. \par
\begin{figure}
 \begin{center}
  \includegraphics[width=1\linewidth, keepaspectratio]{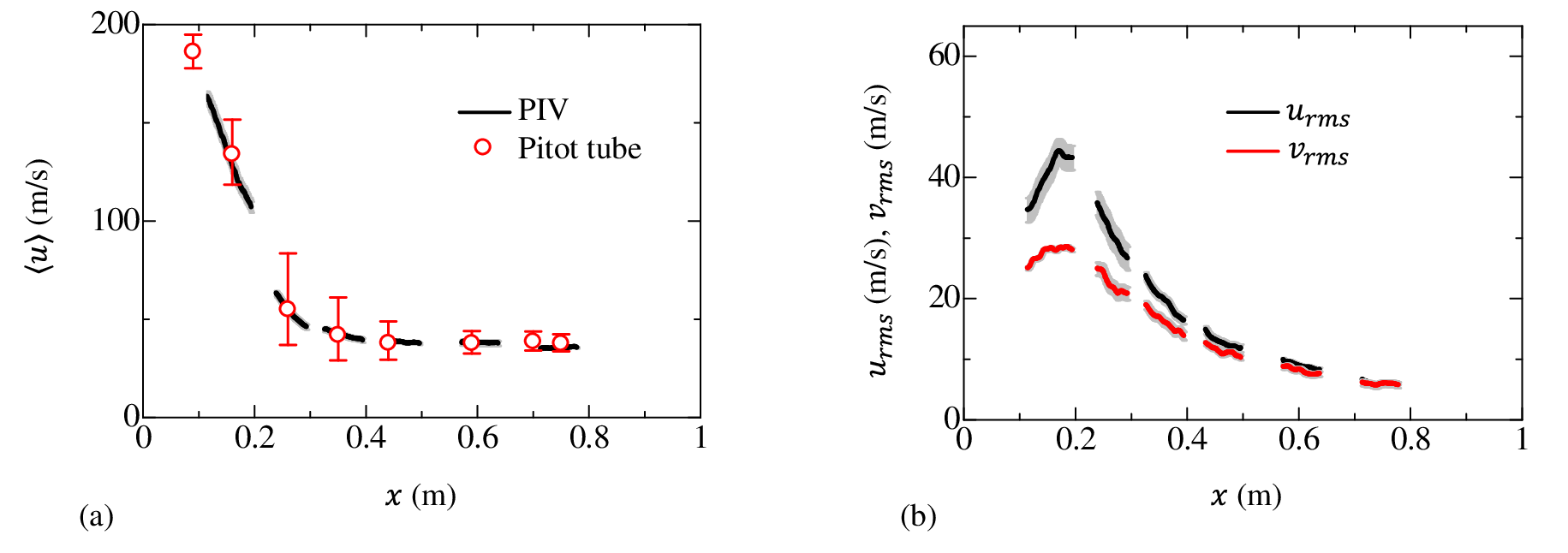}
  \caption{
Streamwise distributions of (a) mean streamwise velocity $\langle u\rangle$ and (b) rms velocity fluctuations of streamwise and vertical velocities, $u_{rms}$ and $v_{rms}$, at $y=0$. The gray shaded area represents the statistical errors estimated by Eq.~($\ref{Eq_Err}$). 
}  
  \label{Fig_Umean_UVrms}
 \end{center}
\end{figure}
Figure~\ref{Fig_Umean_UVrms}(a) shows the streamwise distribution of mean streamwise velocity $\langle u\rangle$. The statistical errors estimated with Eq.~(\ref{Eq_Err}) are shown in gray for PIV data. The mean velocity obtained with the Pitot tube is also plotted for comparison. For the Pitot tube measurement, the error bars represent the mean velocities which correspond to the pressure differences of $\Delta P\pm(\Delta P)_{rms}$,  where $(\Delta P)_{rms}$ is the rms fluctuations of $\Delta P$. For the Pitot tube measurement at $x=0.09$~m, the PIV data for turbulence correction, Eq.~(\ref{Eq_PitotU}), is unavailable. Therefore, an extrapolation of the PIV data provides the velocity variance used for the correction. The mean velocity distribution hardly differs for the PIV and the Pitot tube, ensuring the measurement accuracy of both methods. The mean velocity rapidly decreases in the $x$ direction up to $x\approx0.3$~m and then stays at an almost constant value of 37~m/s. \par
Figure~\ref{Fig_Umean_UVrms}(b) shows the distribution of rms velocity fluctuations of streamwise and vertical velocities, $u_{rms}$ and $v_{rms}$. The rms velocity fluctuations increase with $x$ for $x\lesssim 0.2$~m, where turbulence is generated. They continuously decay in the $x$ direction after the peaks at $x\approx 0.2$~m. Although $u_{rms}$ is much larger than $v_{rms}$ in the upstream region, the difference between $u_{rms}$ and $v_{rms}$ becomes small with $x$. Their ratio $u_{rms}/v_{rms}$ is often used to evaluate the isotropy at large scales. The average of $u_{rms}/v_{rms}$ for $x>0.4~$m is 1.08. This value is close to those in grid turbulence, for which $u_{rms}/v_{rms}\approx 1.1$ has been reported~\citep{krogstad2012near,isaza2014grid,kitamura2014invariants}. Thus, grid turbulence and turbulence generated by the supersonic jet interaction have a similar degree of isotropy. \par
\begin{figure}
 \begin{center}
  \includegraphics[width=1\linewidth, keepaspectratio]{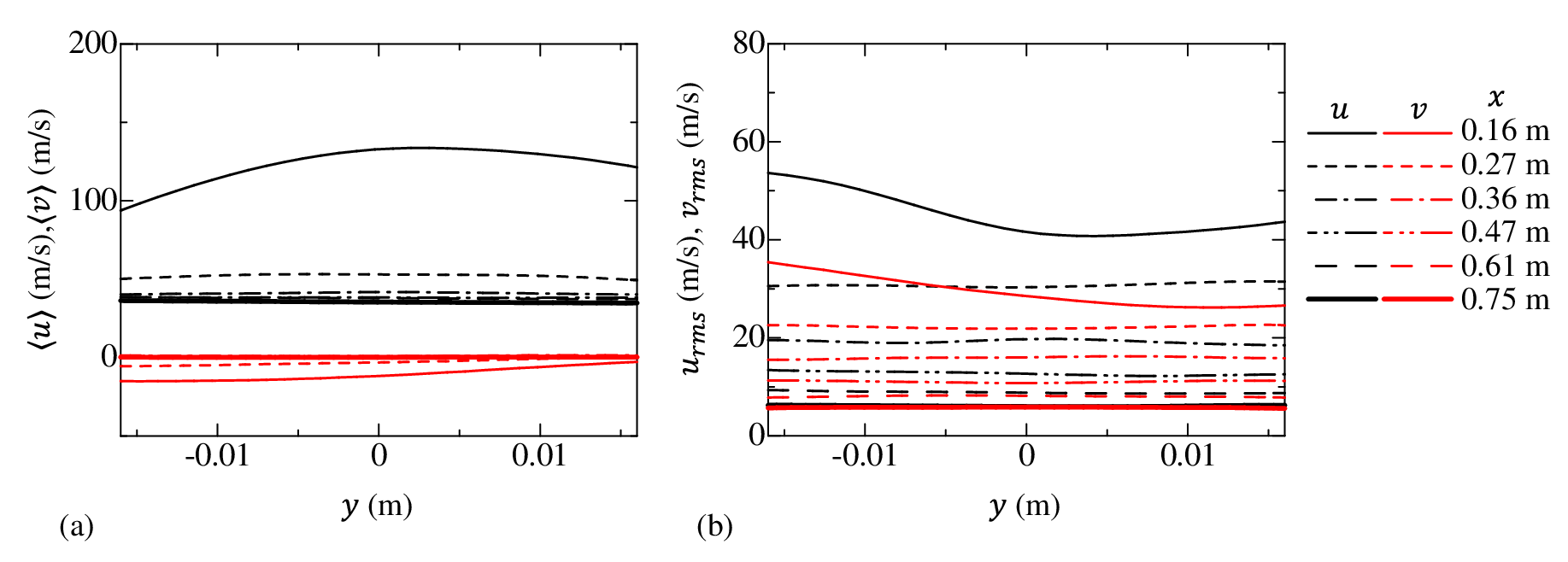}
  \caption{
Vertical distributions of (a) mean velocity and (b) rms velocity fluctuations for streamwise and vertical velocities.  
}  
  \label{Fig_Umean_UVrms_y}
 \end{center}
\end{figure}
Figure~\ref{Fig_Umean_UVrms_y} shows the vertical distributions of mean velocities, $\langle u\rangle$ and $\langle v\rangle$, and rms velocity fluctuations, $u_{rms}$ and $v_{rms}$. At $x=0.16$~m, the mean velocity varies significantly with $y$. The $y$ dependence of the mean velocity becomes weak as $x$ increases. For $x\geq0.36$~m, $\langle u\rangle$ and $\langle v\rangle$ are almost uniform in both $x$ and $y$ directions, and the mean velocity gradient is negligible in the downstream region. The rms velocity fluctuations are also inhomogeneous in the $y$ direction at $x=0.16$~m, where $u_{rms}$ and $v_{rms}$ are large away from the center ($y=0$). Large $u_{rms}$ and $v_{rms}$ at $x=0.16$~m appear in the region of $y$ with the large mean velocity gradient $\partial \langle u\rangle/\partial y$, and the production of turbulent kinetic energy due to the mean shear still influences the flow at $x=0.16$~m. The rms velocity fluctuations become homogeneous in the $y$ direction for $x\geq0.27$~m. These results suggest the nearly homogeneous and isotropic turbulence with a uniform mean flow forms for $x\gtrsim0.3$~m. With the nozzle spacing $L_J=12$~mm and the jet exit diameter $D_E=4.31$~mm, $x\gtrsim0.3$~m corresponds to $x/L_J\gtrsim 25$ and $x/D_E\gtrsim 70$. 
\par
\begin{figure}
 \begin{center}
  \includegraphics[width=1\linewidth, keepaspectratio]{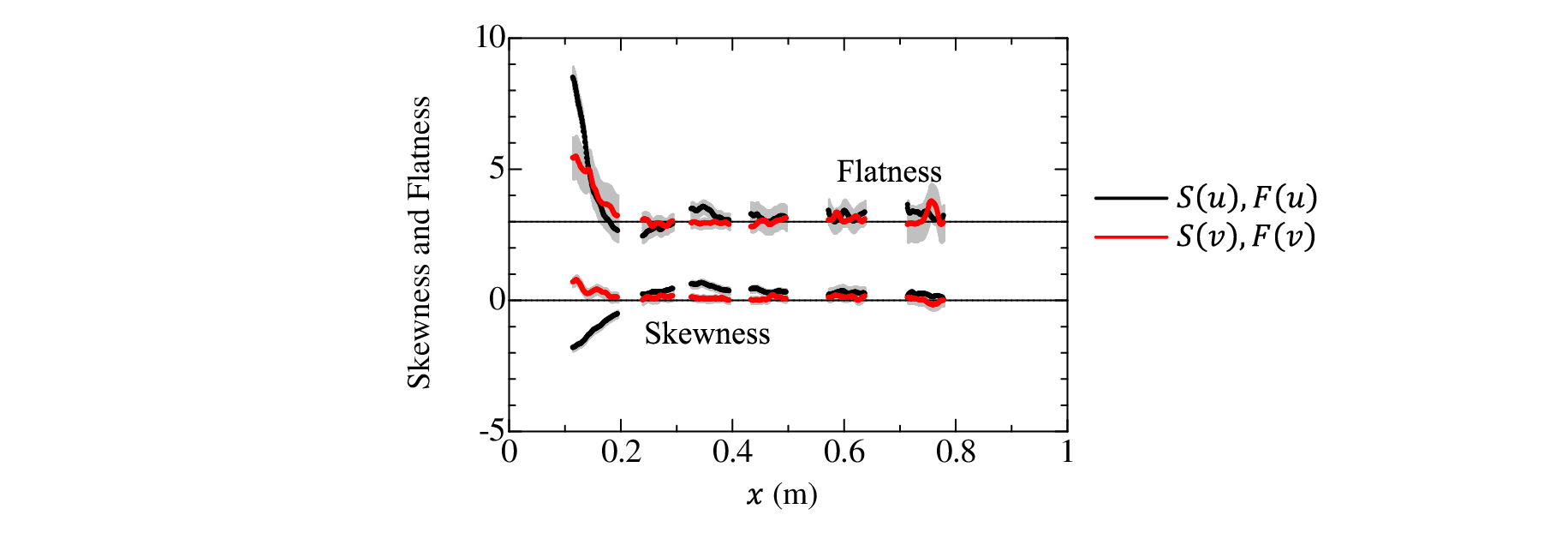}
  \caption{
Streamwise distributions of skewness and flatness of streamwise and vertical velocities at $y=0$. 
}  
  \label{Fig_SkewFlat_U_x}
 \end{center}
\end{figure}
Figure~\ref{Fig_SkewFlat_U_x} shows the streamwise distribution of skewness $S$ and flatness $F$ of velocity components. 
A variable with a probability distribution following a Gaussian function has $S=0$ and $F=3$, shown as horizontal lines in Fig.~\ref{Fig_SkewFlat_U_x}. These Gaussian values of $S$ and $F$ have been obtained for various types of fully-developed isotropic turbulence~\citep{hwang2004creating,krogstad2010grid}. For the present facility, $S$ and $F$ greatly differ from the Gaussian values in the upstream region of $x\lesssim0.15$~m. The flow is highly inhomogeneous and anisotropic in this region. Large $F$ is related to the intermittency of large-scale fluctuations. The intermittent behavior of the jet interaction may cause large values of $F(u)$ and $F(v)$, as also reported for the interaction of two incompressible jets~\citep{zhou2018dual}. As turbulence develops, $S$ and $F$ approach the Gaussian values, and $S\approx0$ and $F\approx 3$ are achieved for $x \gtrsim0.2$. Therefore, the intermittent velocity fluctuations become insignificant once the flow becomes statistically homogeneous in the cross-section. \par
The large-scale turbulent motion is characterized by integral scales, which can be evaluated with velocity auto-correlation functions. The longitudinal auto-correlation functions for streamwise and vertical velocities are respectively defined as 
\begin{align}
  f_u(x,y,r_x)=\frac{\langle u'(x,y)u'(x+r_x,y)\rangle}{\langle u'^{2}(x,y)\rangle},\\
  f_v(x,y,r_y)=\frac{\langle v'(x,y)v'(x,y+r_y)\rangle}{\langle v'^{2}(x,y)\rangle},
\end{align}
where $r_x$ and $r_y$ are separation distances. Because the PIV measures the velocity profiles in a finite area, the available ranges of $r_x$ and $r_y$ are limited. For evaluations of the correlation function with large distances, $f_u(r_x)$ is calculated for $(x,y)$ corresponding to a location near the upstream end of the measurement area at $y=0$. Similarly, $f_v(r_y)$ is calculated for $(x,y)$ corresponding to the upstream bottom corner of the measurement area. In this way, $r_x$ and $r_y$ range up to the size of the measurement area. \par
The longitudinal integral scales are defined as the integrals of $f_u$ and $f_v$: 
\begin{align}
  L_u=\int_0^{r_{x0}} f_udr_x, ~~~
  L_v=\int_0^{r_{y0}} f_vdr_y. 
\end{align}
Here, $r_{x0}$ and $r_{y0}$ are often taken as the first zero-crossing point~\citep{briassulis2001structure}. For the present PIV, the measurement area is insufficient to calculate the correlation functions up to the zero-crossing point. In addition, as also observed for hot-wire measurements in wind tunnels, an accurate evaluation of correlation functions for large $r_x$ and $r_y$ is problematic because of a limited number of large-scale samples. 
For these reasons, the integral scale is often evaluated by approximating the correlation function at large $r$ with an exponential function~\citep{morikawa2008turbulence,bewley2012integral,griffin2019control}. The present study also adapts this methodology for the calculation of integral scales. An exponential function $a_f\mathrm{exp}(-b_{f}r_{x})$ is obtained with a least squares method applied for measured $f_u$ in the range of $r_x$ with $f_u<0.5$. The numerical integration of $f_u$ is calculated with the measured $f_u$ and the exponential function as 
\begin{align}
  L_u=\int_0^{r_{1/2}} f_{u}dr_{x} +\int_{r_{1/2}}^{r_{end}} a_f\mathrm{e}^{-b_{f}r_{x}}dr_x,
\end{align}
where $r_{1/2}$ is defined as $f_u(r_{1/2})=0.5$ and $r_{end}$ corresponds to $r_x$ where $a_f\mathrm{exp}(-b_{f}r_{x})$ is equal to $10^{-5}$. Because of the exponential decay of the correlation, larger values of $r_{end}$ do not affect the evaluation of the integral scale. The integral scale in the vertical direction $L_v$ is also calculated by the same method applied for $f_v$. \par
\begin{figure}
 \begin{center}
  \includegraphics[width=1\linewidth, keepaspectratio]{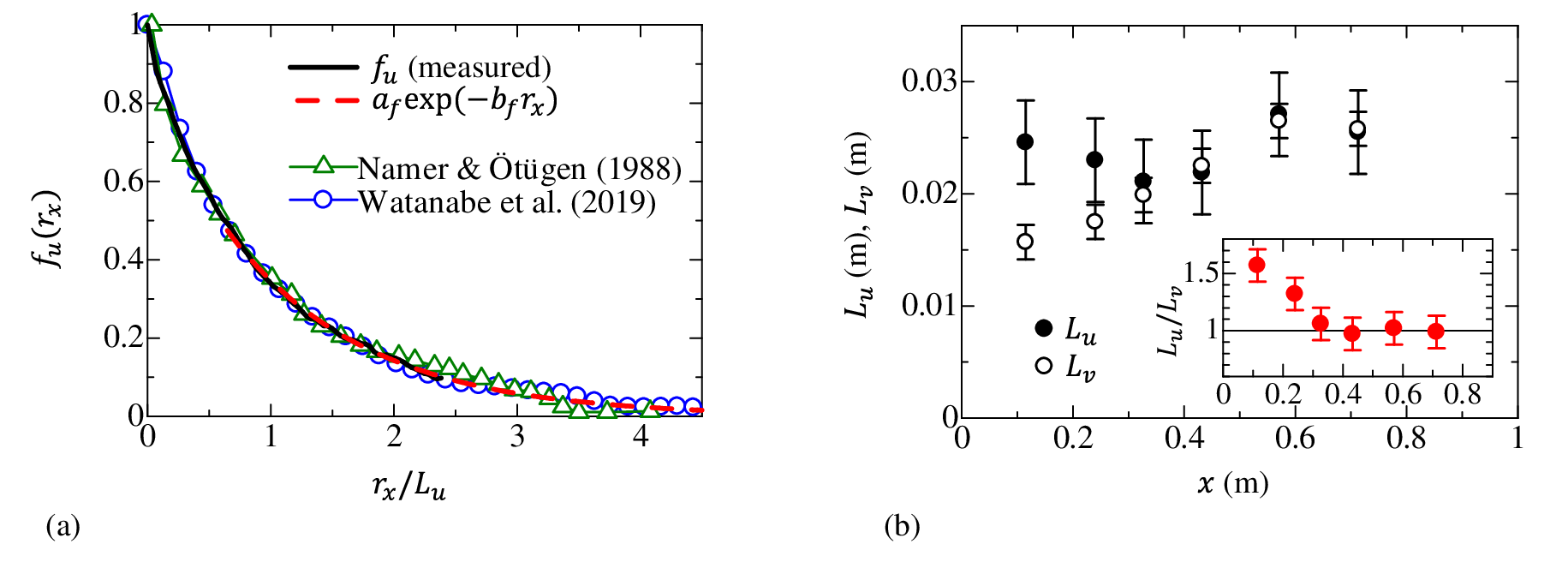}
  \caption{
(a) Longitudinal auto-correlation function of streamwise velocity at $x=570$~mm and $y=0$~mm. The measured correlation is compared with an exponential function $f_u=a_f\mathrm{exp}(-b_{f}r_x)$ and the correlation functions in incompressible turbulent jets (\citealt{namer1988velocity,watanabe2019direct}). A least squares method determines $a_f=0.843$ and $b_{f}=-32.8$~m$^{-1}$ from the measured $f_u$. The separation distance $r_x$ is normalized by the integral scale $L_u$. (b) The streamwise variation of integral scales $L_u$ and $L_v$, which are calculated from the longitudinal auto-correlation functions of streamwise and vertical velocities, respectively. The inset shows the ratio between $L_u$ and $L_v$. 
}
  \label{Fig_Luv}
 \end{center}
\end{figure}
Figure~\ref{Fig_Luv}(a) shows the correlaton function $f_u$ and its approximation based on the exponential function at $(x,y)=(570$~mm$, 0$~mm$)$. The separation distance $r_x$ is normalized with the integral scale for comparisons with other turbulent flows. The exponential function accurately approximates the decay of $f_u$ for large $r_x$. In addition, the profile of $f_u$ agrees well with the correlation functions measured in other turbulent flows. \par
Figure~\ref{Fig_Luv}(b) presents the streamwise variation of the integral scales. In the inhomogeneous and anisotropic region, $x<0.3$~m, the streamwise integral scale $L_u$ is larger than the vertical scale $L_v$, and large-scale velocity fluctuations are also anisotropic. On the other hand, the downstream region satisfies $L_u\approx L_v$. This is also confirmed by the inset presenting $L_u/L_v$. The average of $L_u/L_v$ for $x>0.3$~m is 1.03, and the integral scales hardly differ in both directions, suggesting that turbulent motion at large scales is statistically isotropic. For $0.3 \lesssim x \lesssim 0.6$, the integral scales increase as turbulence decays. However, in the most downstream location of the measurement, $L_u$ and $L_v$ decrease with $x$. In DNS of decaying isotropic turbulence in a triply periodic domain, the decay is influenced by the boundary conditions when the domain size divided by the integral scale is smaller than about 3~\citep{anas2020freely}. The side length of the cross-section, 0.1~m, is about $3.7L_u$ at $x=0.57$~m. The decrease of $L_u$ and $L_v$ in the further downstream region may be caused by the confinement effects. \par
\begin{figure}
 \begin{center}
  \includegraphics[width=1\linewidth, keepaspectratio]{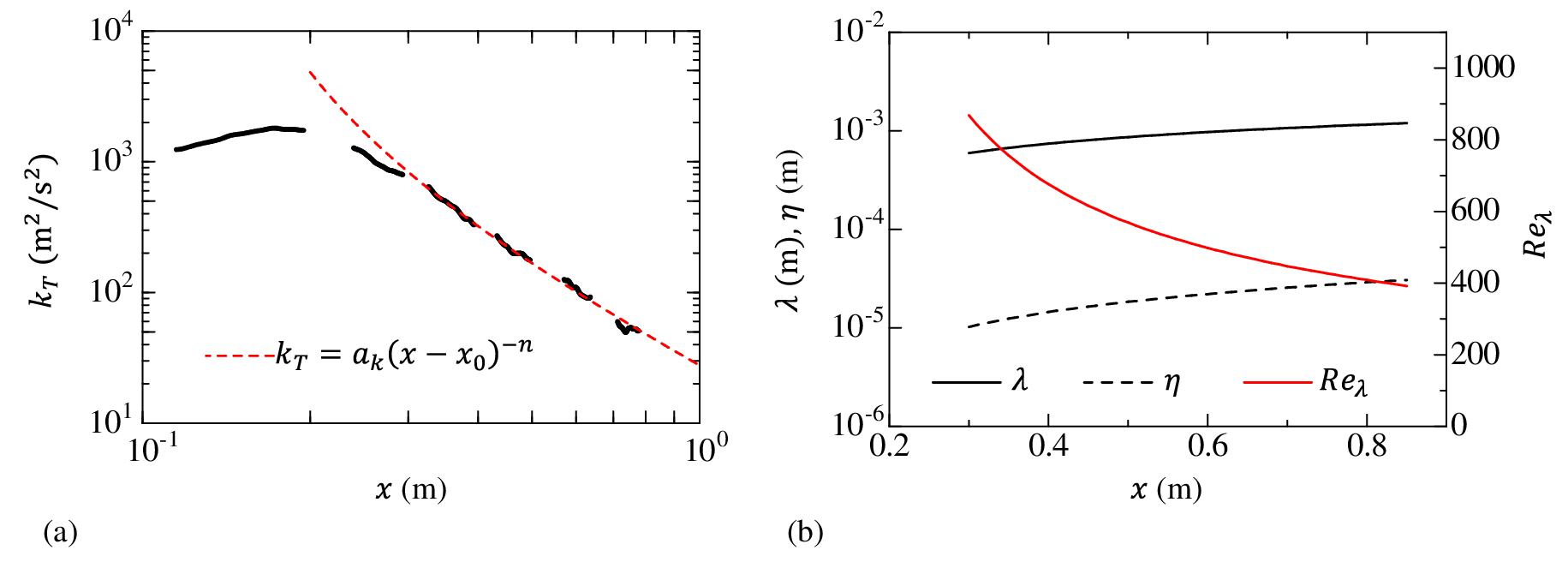}
  \caption{
(a) The decay of turbulent kinetic energy $k_T$ along the centerline. The red broken line represents a power law $k_T=a_k(x-x_0)^{-n}$ with $a_k=21.0$ m$^2$/s$^2$, $x_0=0.12$~m, and $n=2.1$, which are obtained with a least squares method applied for $x>0.32$~m. (b) The streamwise variations of Taylor microscale $\lambda$, Kolmogorov scale $\eta$, and turbulent Reynolds number $Re_\lambda$, estimated with the power law decay of $k_T$. 
}  
  \label{Fig_TKE}
 \end{center}
\end{figure}
\subsection{The dissipation rate of turbulent kinetic energy}\label{sec_dis}
Figure~\ref{Fig_TKE}(a) shows the streamwise variation of turbulent kinetic energy per unit mass $k_T\approx(\langle {u'}^2\rangle+2\langle {v'}^2\rangle)/2$ along the centerline. The decay of $k_T$ is investigated for the nearly homogeneous and isotropic region, $x>0.32$~m. A nonlinear least squares method (Levenberg--Marquardt method) applied for $x>0.32$~m yields the parameters in the power law, Eq.~(\ref{Eq_decay}), as $a_k=21.0$ m$^2$/s$^2$, $x_0=0.12$~m, and $n=2.1$. Equation~(\ref{Eq_decay}) with these parameters is also shown in the figure. The decay of $k_T$ is well approximated by the power law. The exponent $n=2.1$ is larger than typical values reported for incompressible grid turbulence at high Reynolds numbers, $n\approx1.0$--1.5~\citep{kistler1966grid,kang2003decaying,thormann2014decay,zheng2021turbulent}. As discussed in Sect.~\ref{Sec_Introduction}, \cite{zwart1997grid} also reported that $n$ in compressible grid turbulence tends to be larger than incompressible cases. This tendency is qualitatively consistent with the large exponent in this study. However, these results contradict the decrease of $n$ observed in shock-tube experiments~\citep{briassulis2001structure,fukushima2021impacts}, and the compressibility effects on the decay exponent should be addressed in future studies. \par
The small-scale turbulence properties are often related to the kinetic energy dissipation rate. The Taylor microscale $\lambda$ and Kolmogorov scale $\eta$ are defined as 
\begin{align}
\lambda&=\sqrt{\frac{10\mu k_T}{\rho \varepsilon}},
  \label{Eq_lambda}\\
\eta&=(\mu/\rho)^{3/4}\varepsilon^{-1/4}.
  \label{Eq_eta}
\end{align}
The turbulent Reynolds number $Re_\lambda$ is define with $\lambda$ as 
\begin{align}
Re_\lambda=\frac{\rho \sqrt{2k_T/3}\lambda}{\mu}. 
  \label{Eq_Rel}
\end{align}
The dissipation rate estimated from the decay of $k_T$ with Eq.~(\ref{Eq_Epsilon2}) is used to evaluate $\lambda$, $\eta$, and $Re_\lambda$. Figure~\ref{Fig_TKE}(b) shows their streamwise variations in the nearly homogeneous and isotropic region. $\lambda$ and $\eta$ increase as turbulence decays in the $x$ direction. The orders of the length scales are 
$\lambda=O(10^{-3}$~m$)$ and
$\eta=O(10^{-5}$~m$)$. The spatial resolution of PIV is about $1.0\times10^{-3}$~m, and the scales greater than the Taylor microscale are resolved by the PIV. This resolution is sufficient to investigate the large-scale properties, such as mean velocity and rms velocity fluctuations. The turbulent Reynolds number decreases with the decay of turbulence. It ranges between about 800 and 400. These values of $Re_\lambda$ are larger than those of most subsonic grid turbulence in large wind tunnel facilities because of the large velocity fluctuations of supersonic jets. \par
\begin{figure}
 \begin{center}
  \includegraphics[width=1\linewidth, keepaspectratio]{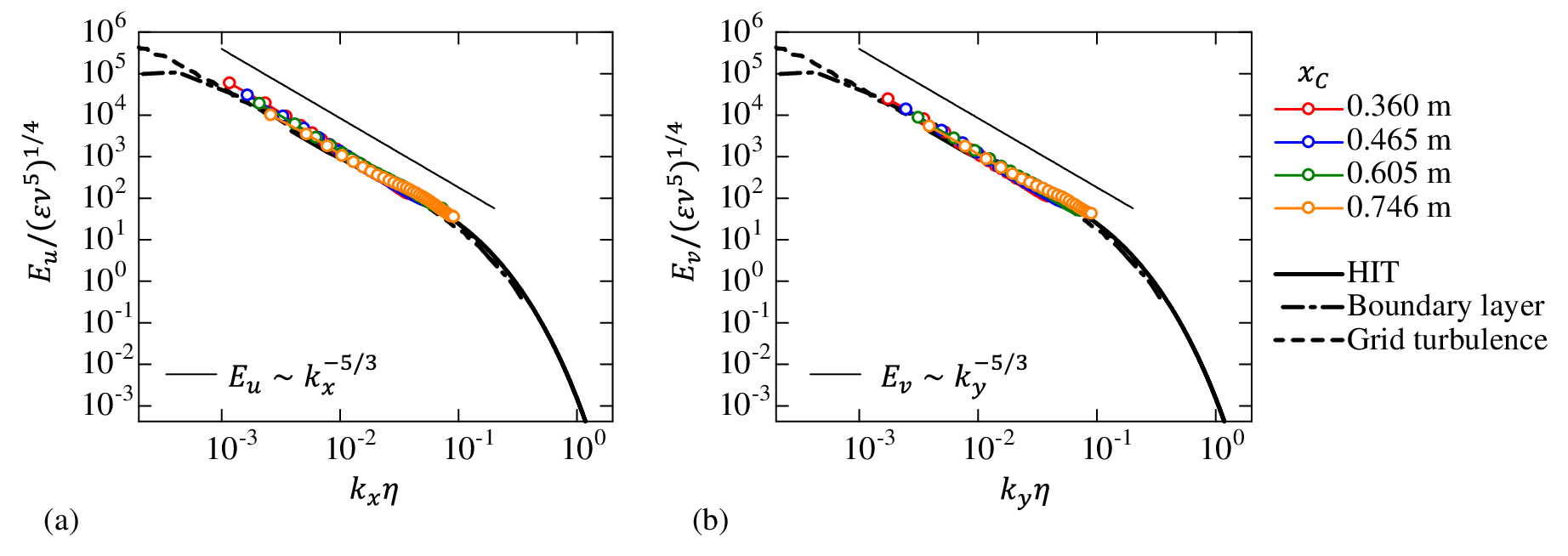}
  \caption{
Normalized energy spectra of (a) streamwise velocity, $E_u(k_x)$, and (b) vertical velocity, $E_v(k_y)$, where $k_x$ and $k_y$ are the wavenumbers in the streamwise and vertical directions, respectively. The spectra and wavenumbers are non-dimensionalized with the kinematic viscosity $\nu=\mu/\rho$, the energy dissipation rate per unit mass $\varepsilon$, and the Kolmogorov scale $\eta$. The present results are compared with direct numerical simulation of forced HIT ($Re_\lambda=202$)~\citep{watanabe2023response} and experiments of a boundary layer ($Re_\lambda=202$)~\citep{saddoughi1994local,nieuwstadt2016introduction} and grid turbulence ($Re_\lambda=520$)~\citep{kistler1966grid}.
}  
  \label{Fig_EuEv}
 \end{center}
\end{figure}
One-dimensional energy spectra of streamwise and vertical velocities ($E_u$ and $E_v$) are calculated with Fourier transform in the $x$ and $y$ directions, respectively. The wavenumbers in these directions are denoted by $k_x$ and $k_y$. The energy spectra in turbulence with a high Reynolds number are known to have a universal shape in an inertial subrange in various turbulent flows~\citep{pope2000turbulent}. Figure~\ref{Fig_EuEv} presents $E_u$ and $E_v$ measured at several streamwise locations. The spectra and wavenumbers are normalized by the dissipation rate $\varepsilon$ and kinematic viscosity $\nu=\mu/\rho$. The present results are compared with incompressible turbulent flows. The spectra are evaluated for normalized wavenumbers between $10^{-3}$ and $10^{-1}$, determined by the measurement area and spatial resolution. The present results quantitatively agree with other turbulent flows. They are also consistent with the power laws expected for the inertial subrange, $E_u\sim k_x^{-5/3}$ and $E_v\sim k_y^{-5/3}$. Thus, the present PIV with condensation particles is capable of measuring the scale dependence of velocity fluctuations in the inertial subrange. In addition, the flow at a wide range of scales is in a statistically isotropic state, as attested by $E_u\approx E_v$. \par
\begin{figure}
 \begin{center}
  \includegraphics[width=1\linewidth, keepaspectratio]{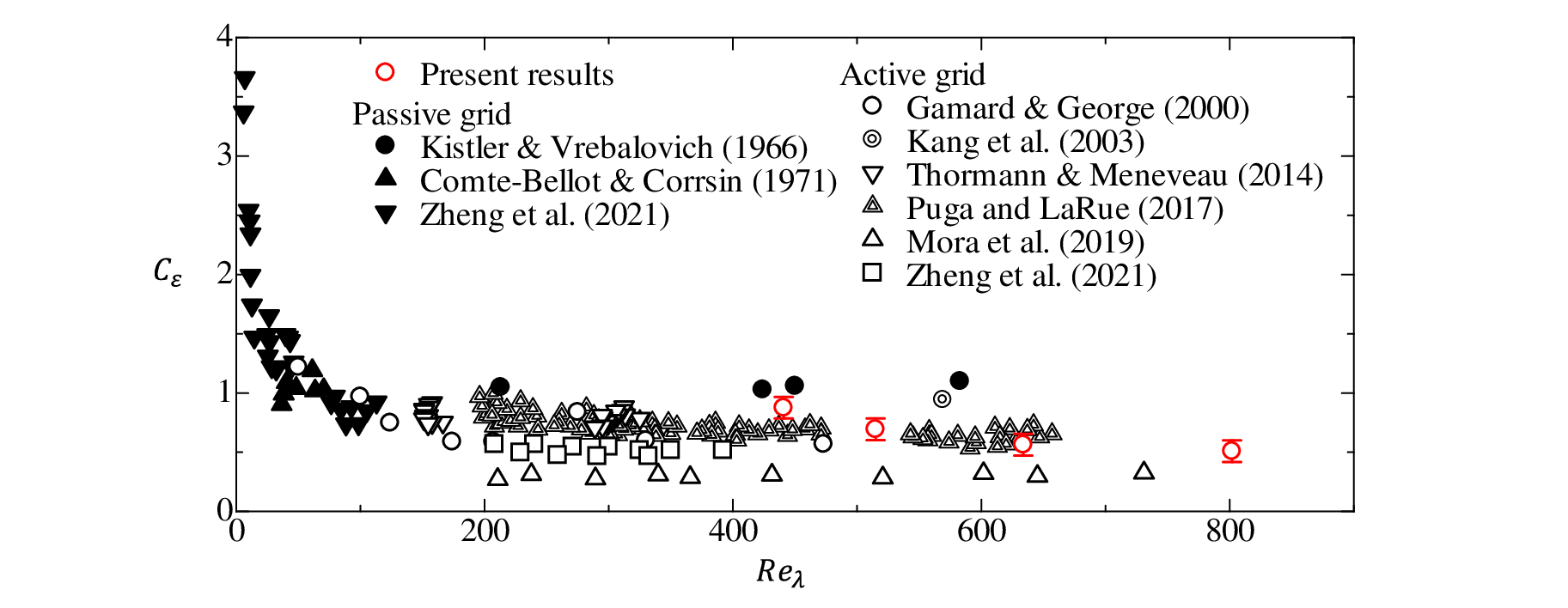}
  \caption{
Dependence of non-dimensional energy dissipation rate $C_\varepsilon$ on the turbulent Reynolds number $Re_\lambda$. The present results are compared with experimental data of subsonic grid turbulence: 
\cite{kistler1966grid},
\cite{comte1971simple}, and 
\cite{zheng2021energy} for passive-grid generated turbulence; 
\cite{gamard2000reynolds},
\cite{kang2003decaying},
\cite{thormann2014decay},
\cite{puga2017normalized},
\cite{mora2019energy}, and 
\cite{zheng2021turbulent} for active-grid generated turbulence.
}
  \label{Fig_Cepsi}
 \end{center}
\end{figure}
The dissipation rate is often related to characteristic velocity ${\cal U}$ and lenght ${\cal L}$ of large-scale turbulent motion as 
\begin{align}
\varepsilon=C_\varepsilon\frac{{\cal U}^3}{\cal L},
  \label{Eq_Cepsilon}
\end{align}
where $C_\varepsilon$ is a non-dimensional dissipation rate. $C_\varepsilon$ plays a central role in theories and models of turbulent flows~\citep{vassilicos2015dissipation}. 
The velocity and length scales in the present experiment are evaluated as ${\cal U}=\sqrt{(\langle u'^2\rangle+\langle v'^2\rangle)/2}$ and ${\cal L}=(L_u+L_v)/2$, and $C_\varepsilon$ is evaluated with $\varepsilon$ in the nearly homogeneous and isotropic region. Figure~\ref{Fig_Cepsi} plots $C_\varepsilon$ as a function of $Re_\lambda$ for the present experiment and other studies of incompressible grid turbulence. In fully-developed grid turbulence with high $Re_\lambda$ in incompressible flows, $C_\varepsilon$ takes values approximately between 0.4 and 1.1 while $C_\varepsilon$ increases rapidly as $Re_\lambda$ becomes smaller than about 100. The present values of $C_\varepsilon=0.51$--0.87 are within a range reported for grid turbulence with high $Re_\lambda$. The decay exponent of the turbulent kinetic energy is known to be influenced by the spatial variation of $C_\varepsilon$. For the present experiment, $C_\varepsilon$ slightly increases in the streamwise direction. The increase of $C_\varepsilon$ causes a larger decay exponent $n$~\citep{krogstad2010grid}. The spatial variation of $C_\varepsilon$ is possibly relevant to the large values of $n$ in the present experiment and in compressible grid turbulence~\citep{zwart1997grid}. \par
\section{Conclusion}\label{Sec_Conclusions}
The present study has reported the development and characterization of the multiple-supersonic-jet wind tunnel, which is designed to investigate decaying nearly homogeneous and isotropic turbulence in a compressible flow regime. The PIV, pitot tube, and thermocouple measurements characterize the generated turbulence, where velocity vectors, static and dynamic pressures, and temperature are measured. The present wind tunnel can sustain a steady mean flow over 2--3~s. It can also seed tracer particles in the test section by condensation. The accuracy of the PIV is confirmed by comparing the velocity statistics with other turbulent flows. In addition, the mean velocity measured by the PIV agrees well with the Pitot tube measurements, for which turbulence correction is applied with the PIV data of velocity variances. \par
The mean velocity in the upstream region is large along the centerline of the test section. However, as the flow evolves, the mean velocity profile becomes uniform in the test section. Then, the mean velocity gradient becomes negligibly small in the downstream region. Once the mean flow profile becomes uniform, nearly homogeneous and isotropic turbulence begins to decay. At this decay regime, rms velocity fluctuations are also uniform in the cross-section. The ratio of rms values of streamwise- and vertical-velocity fluctuations is about 1.08, close to the values reported for incompressible grid turbulence. The longitudinal integral length scales are also similar in the streamwise and vertical directions. These statistics have confirmed that the supersonic jet interaction generates nearly homogeneous and isotropic turbulence. In the inhomogeneous and anisotropic region near the jet nozzles, the flatness of velocity fluctuations is larger than 3, indicating that the large-scale velocity fluctuations are highly intermittent. The skewness and flatness of velocity fluctuations are close to the Gaussian values in the decay region, where the large-scale intermittency is insignificant. \par
The decay of turbulent kinetic energy per unit mass is investigated in the nearly homogeneous and isotropic region. The turbulent kinetic energy decays according to a power law, as reported for subsonic grid turbulence. The decay exponent $n$ for the supersonic jet interaction is about 2, which is larger than the values reported for incompressible grid turbulence. This large value of $n$ is consistent with experiments of compressible grid turbulence in a high-speed wind tunnel by \cite{zwart1997grid}. In addition, the dissipation rate of turbulent kinetic energy per unit mass has been evaluated with the decay of the turbulent kinetic energy. The energy spectra of velocity fluctuations normalized by the dissipation rate and kinematic viscosity follow the universal power law of the inertial subrange, observed for other incompressible turbulent flows. The non-dimensional dissipation rate $C_\varepsilon$ for the supersonic jet interaction is within a range of 0.51--0.87, which is also consistent with grid turbulence at a high Reynolds number. 
\par
The nozzle component in the present wind tunnel can be replaced with other components with different nozzle geometries, generating turbulent jets with different Mach and Reynolds numbers. This feature is helpful in varying the degree of the compressibility effects, which will be considered in future studies. These results have demonstrated that the multiple-supersonic-jet wind tunnel is helpful in the investigation of decaying homogeneous isotropic turbulence in a compressible flow regime. \par

\section*{Statements and Declarations}
\subsection*{Funding}
This work was supported by Japan Society for the Promotion of Science (KAKENHI Grant Numbers JP22K03903 and JP22H01398).

\subsection*{Competing interests}
The authors have no competing interests to declare that are relevant to the content of this article.

\subsection*{Availability of data and materials}
The data that support the findings of this study are available from the corresponding author upon reasonable request.

\subsection*{Authors' contributions}
{\bf{Takahiro Mori}}: 
Data curation (equal);
Formal Analysis (equal);
Investigation (lead);
Methodology (equal);
Resources (equal);
Software (supporting);
Validation (equal);
Visualization (equal);
Writing - original draft (supporting);
Writing - review \& editing (equal).
{\bf{Tomoaki Watanabe}}:
Conceptualization (lead);
Data curation (equal);
Formal Analysis (equal);
Funding acquisition (equal);
Investigation (supporting);
Methodology (equal);
Project administration (lead);
Resources (equal);
Software (lead);
Supervision (lead);
Validation (equal);
Visualization (equal);
Writing - original draft (lead);
Writing - review \& editing (equal).
{\bf{Koji Nagata}}:
Conceptualization (supporting);
Data curation (supporting);
Funding acquisition (equal);
Investigation (supporting);
Methodology (supporting);
Project administration (supporting);
Resources (equal);
Software (supporting);
Supervision (supporting);
Validation (equal);
Visualization (supporting);
Writing - original draft (supporting);
Writing - review \& editing  (equal).

\bmhead{Acknowledgments}
The authors acknowledge Prof. K. Mori (Osaka Prefecture University), Mr. N. Iwakura (Nagoya University), and Ms. R. Nakayama (Nagoya University) for their help and valuable comments in developing the multiple-supersonic-jet wind tunnel. The authors also acknowledge Mr. K. Ishizawa (Nagoya University) for assistance in experiments. 


\end{document}